\documentclass[secnumarabic, reprint, superscriptaddress, tightenlines, amsmath,amssymb, aps, prl, nobibnotes,
]{revtex4-1}

\usepackage{blindtext}
\usepackage{float}
\usepackage{tabularx}
\usepackage{adjustbox}      
\usepackage{booktabs}
\usepackage{natbib}
\usepackage{graphicx,caption,subcaption,xcolor}
\usepackage[colorlinks=true,citecolor=blue,
linkcolor=blue,urlcolor=blue]{hyperref}
\usepackage[utf8]{inputenc} 
\usepackage[english]{babel}
\usepackage[OT2, T1]{fontenc}
\usepackage{upquote}
\usepackage{mathtools}
\usepackage{bigints}

\begin{document}

\title{
Many-body effects of two-level systems in superconducting qubits
%a Josephson junction qubit
}

\author{Joshuah T. Heath}
\email{joshuah.t.heath@su.se}
\author{Alexander C. Tyner}

\affiliation{Nordita, Stockholm University and KTH Royal Institute of Technology, Hannes Alfvéns väg 12, SE-106 91 Stockholm, Sweden}
\affiliation{Department of Physics, University of Connecticut, Storrs, Connecticut 06269, USA}

\author{Thue Christian Thann}

\affiliation{NNF Quantum Computing Programme, Niels Bohr Institute, University of Copenhagen, Denmark}

\author{\phantom{.}\\ Vincent P. Michal}

\affiliation{NNF Quantum Computing Programme, Niels Bohr Institute, University of Copenhagen, Denmark}

\author{Peter Krogstrup}

\affiliation{NNF Quantum Computing Programme, Niels Bohr Institute, University of Copenhagen, Denmark}

\author{Mark Kamper Svendsen}

\affiliation{NNF Quantum Computing Programme, Niels Bohr Institute, University of Copenhagen, Denmark}

%\affiliation{Max Planck Institute for the Structure and Dynamics of Matter, Center for Free-Electron Laser Science (CFEL), Luruper Chaussee 149, 22761 Hamburg, Germany}

\author{Alexander V. Balatsky}
\email{balatsky@kth.se}
\affiliation{Nordita, Stockholm University and KTH Royal Institute of Technology, Hannes Alfvéns väg 12, SE-106 91 Stockholm, Sweden}
\affiliation{Department of Physics, University of Connecticut, Storrs, Connecticut 06269, USA}

\date{\today}

\begin{abstract}
\noindent 
%Josephson junctions are a key component for the construction of quantum bits, as the low dissipation inherent to superconducting media results in long coherence times for the qubit. 
Superconducting qubits are often adversely affected by two-level systems (TLSs) within the Josephson junction, which contribute to decoherence and subsequently limit the performance of the qubit. By treating the TLS as a soft (i.e., low-frequency) bosonic mode localized in real space, we find that a single TLS in either the amorphous oxide surface or the superconducting bulk may result in a localized "hot spot" of amplified Josephson energy. Such amplification is shown to have a non-negligible effect on the $T_1$ time of certain superconducting qubits, regardless of whether or not the TLS is on resonance with the qubit frequency. With this study, we identify sources of decoherence unique to the superconducting element of superconducting quantum devices. 
\end{abstract}

\pacs{1}

\maketitle

\section{I. Introduction}
Contemporary solid state qubits are often plagued by parasitic two-level systems (TLSs) that result in a degradation of device quality~\cite{Simmonds2004Aug,Muller2019Oct}. If even a single TLS is coupled resonantly to the Josephson junction, then decoherence may occur when the TLS decays from its excited state~\cite{Ku2005Jul,Martinis2005Nov}. %Likewise, an ensemble of such TLSs has been proposed to be a major source of $1/f$ noise~\cite{Black1983Oct,Burnett2014Jun}.
Consequently, several mitigation strategies have been proposed to reduce the detrimental influence of TLSs on qubit performance, including the consideration of Josephson junctions with smaller area~\cite{Martinis2005Nov} and the construction of amorphous oxide tunnel junctions from materials of relatively large 
atomic mass~\cite{Place2021Mar,Wang2024May,Wang2022Jan,Tyner2025}.

While TLSs are often ubiquitous in the oxide tunnel junctions of superconducting quantum devices, a comprehensive microscopic explanation is still highly contested~\cite{Martinis2005Nov,DuBois2013Feb,Agarwal2013Apr,Gordon2014Dec,Zanker2016Jan,Muller2019Oct, deGraaf2020Dec}. Regardless of the precise microscopic origin of TLSs, most proposals share fundamental properties of the standard tunneling model~\cite{W.Anderson1972Jan,Phillips1972May,BibEntry1981}, which treats the TLS as a local anharmonic double-well potential. Such local anharmonic potentials result in localized soft phonons with frequencies well below the boson peak characteristic of disordered materials~\cite{Wang2019Jan}. The association of TLSs with localized low-frequency phonons is likewise supported by recent {\it ab initio} simulations carried out on amorphous Al$_2$O$_3$ ($\alpha$-Al$_2$O$_3$) and Ta$_2$O$_3$~\cite{Tyner2025}. 
\begin{figure}[t!]
%\vspace{7mm}
 \raggedright
\hspace{-5.0mm}\includegraphics[width=\dimexpr1.4\columnwidth-3cm]{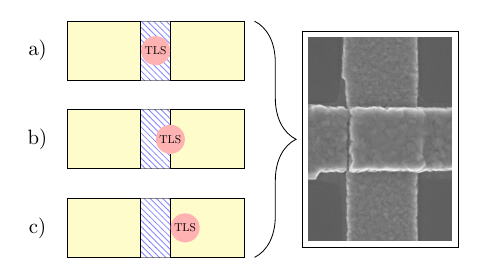}
\vspace{0mm} \caption{\small We consider the TLS in three separate regions of the Josephson junction (right): a) TLS in the normal amorphous oxide (blue hatched); b) TLS at the interface of the normal oxide and the superconductor (yellow); and c) TLS in the superconductor. SEM picture of Al Josephson junction courtesy of NNF Quantum Computing Programme.}
\label{Fig1}
\end{figure}
 \begin{center}
\begin{table*}[t!]
\centering
{\begin{tabular}{| c| c | c | c | c | c || }
  \hline
 & \phantom{\LARGE D}\hspace{-5mm} \,Material\, & \, bare e-p coupling ($\lambda^{(0)}$)  \,& \,$\overline{\Delta(\infty,\,\omega)/\omega_E}$ \, & \,$\overline{\Delta(0,\,\omega)/\omega_E}$ \, & \, $\overline{\Delta(0,\,\omega)/\Delta(\infty,\,\omega)}$ \, \\ 
  \hline
   & Al & 0.43 & 0.0009 & 0.0002 & 0.9544  
\\  \hline
   & Ta & 0.69 & 0.0032 & 0.0007 & 0.9422 \\ 
   \hline
      & Nb & 1.11 & 0.0084 & 0.0016 & 0.8976 \\ 
   \hline
\end{tabular}}
\caption{\small Electron-phonon (e-p) coupling strength and averages of the gap function for Al, Ta, and Nb found via Eliashberg theory. The e-p coupling is reported as it's "bare" value; i.e., the value of the coupling strength in a crystalline sample of the material in question, as reported in Ref.~\cite{RevModPhys.62.1027}. As $\lambda^{(0)}$ increases, the average of the gap function over the full frequency range in the absence of the TLS (i.e., $\overline{\Delta(\infty,\,\omega)/\omega_E}$) increases. The reduction of the average gap function at the site of the TLS ($\overline{\Delta(0,\,\omega)/\omega_E}$) also increases with increasing $\lambda^{(0)}$.} 
\end{table*} 
\end{center}
\vspace{-14mm}

\indent We claim that viewing the TLS as a localized soft phonon reveals a new low-frequency channel for decoherence beyond resonant TLS coupling to the qubit. TLSs typically live either in the tunnel barrier~\cite{DuBois2013Feb,Gordon2014Dec,Zeng2016Jul} or the interface of the amorphous oxide and the superconductor, where oxygen vacancies result in gapless interface states~\cite{Jung2009Sep,Zeng2016Jul,Tyner2025}. If we treat the TLS as a localized bosonic mode, the presence of a single TLS opens a new inelastic scattering channel for free electrons on the surface of the oxide~\cite{Marsiglio1991Sep,Ziman2001,Chubukov2020Jun,Heath2024Jul}. This new scattering channel leads to a change in the local density of states (LDOS) which, in turn, leads to a change in the local normal-state conductivity~\cite{Stipe1998Jun,Zhu2004Jul,Balatsky2006May,Behm2010Dec} and a local modulation of the Josephson energy~\cite{PhysRevLett.10.486,Fulton1968Nov}. 
%As a result, we expect localized "hot spots" of amplified Josephson energy centered around TLSs at the interface. In a current-biased Josephson junction (i.e., phase qubit), this will result in a reduction of the $T_1$ time~\cite{Martinis2003Mar,Martinis2005Nov}. 
Likewise, %it is important to note that
%, while TLSs are often considered to live at the amorphous interface of a Josephson junction, 
there is now significant evidence for TLSs to be present in the bulk superconductor, primarily due 
%Such TLSs may be due 
to spatial fluctuations of the superconducting order parameter~\cite{Bespalov2016Sep,deGraaf2020Dec}, magnetic defects within the superconductor~\cite{Huang2024Dec}, or the granular nature of the superconductor itself~\cite{Grunhaupt2019Aug,Kristen2024May}. If we view the TLS as a soft phonon, then a TLS in the superconductor may influence the local structure of the many-body gap function~\cite{Millis1988Apr,Morr2003Aug,Balatsky2003Dec,Balatsky2006May}, thereby leading to potential decoherence mechanisms unique to the superconducting nature of the state. 
%Indeed, STM spectra has already suggested in-gap, low-frequency bosonic modes in granular Al films which lead to a modulation of the superconducting LDOS~\cite{Yang2020Sep}.

%The soft phonon picture of the TLS motivates us to present a comparative study of TLSs localized in the amorphous oxide interface and superconducting bulk. Specifically, w
Taking the soft phonon picture of the TLS, we consider the influence of {\it a single TLS} on the normal-state conductivity and the superconducting gap. We then study how a TLS will modulate the local Josephson energy and, subsequently, the decoherence of a simple superconducting qubit. We do this through the lens of many-body perturbation theory, borrowing theoretical tools from inelastic electron tunnelling spectroscopy (IETS)~\cite{Stipe1998Jun,Balatsky2003Dec,Zhu2004Jul,Balatsky2006May,Behm2010Dec} and the theoretical and numerical machinery of Eliashberg theory~\cite{Eliashberg1, Eliashberg2, RickayzenBook, Parks1, Bardeen1973Jul,RevModPhys.62.1027,Marsiglio2020Jun,Chubukov2020Jun} so that we may account for the material-specific parameters of Al, Ta, and Nb while simultaneously treating the TLS as a localized {\it dynamical impurity}~\cite{Balatsky2006May}.
%. In particular, the latter allows us to conduct a material-dependent study of aluminium (Al), tantalum (Ta), and niobium (Nb) devices by considering the electron-phonon interaction strength unique to the compound in question. 
We ultimately find that a single TLS will strongly affect the performance of smaller current-biased Josephson junctions constructed from materials with larger electron-phonon coupling, such as Nb. 
% materials with larger electron-phonon coupling 

%\begin{widetext}

%\end{widetext}

% Our study illustrates how a single TLS will adversely affect the performance of a current-biased Josephson junction, and how such effects are dependent upon the junction's overall size and the electron-phonon coupling strength intrinsic to the constituent material.

% We find that the local electron-phonon coupling is amplified near the low-frequency TLS, which leads to a much stronger amplification of the local Josephson energy for Nb devices as compared to Al or Ta. 

%Our study thus provides a "materials roadmap" for the mitigation of off-resonant decoherence channels vis-à-vis two-level systems localized throughout the entirety of the junction.

\section{II. Two-level systems in the amorphous oxide junction and junction/superconductor interface}

The current through a Josephson junction is given by $I(t)=I_c \sin \delta$, where $\delta$ is the phase difference of the Ginzburg–Landau order parameters across the junction~\cite{Josephson1962Jul,Josephson1974Apr} and $I_c$ is the DC critical current, given by~\cite{PhysRevLett.10.486,Fulton1968Nov}

\begin{align}
    I_c = \dfrac{2}{\pi e R_N}\int_0^\infty d\omega\,\bigg[\Re\left\{\dfrac{\Delta(\omega)}{\sqrt{\omega^2-\Delta^2(\omega)}}\right\}\bigg]^2\label{AB},
\end{align}

\noindent where $\Delta(\omega)$ is the gap function of the superconductor,
%~\footnote{Note that, in the case of weak electron-phonon coupling (e.g. Al), the details of the phonon dynamics become less important, and to a good approximation~\cite{Marsiglio2018Jul} we may take the gap function $\Delta(\omega)\approx \Delta_0 $ to be purely real and frequency independent, in which case we obtain the more well-known Ambegaokar–Baratoff relation $I_0=\pi \Delta_0/(2eR_N)$~\cite{PhysRevLett.10.486}.
%For materials with stronger electron-phonon coupling (e.g. Ta or Nb), this approximation is no longer appropriate, and Eqn.~\eqref{AB} must be used in full.},
$R_N$ is the normal-state resistance of the junction, $e$ is the electron charge, and the temperature $1/\beta$ is taken appreciably close to zero. The characteristic parameter of the Josephson junction (the Josephson energy) is defined in terms of the critical current as $\epsilon_J\equiv \Phi_0 I_c/(2\pi)$, where $\Phi_0=h/(2e)$ is the flux quantum. Note that the specific frequency dependence of the gap function $\Delta(\omega)$ is determined by dynamical phonon effects inherent to the superconducting material~\cite{RevModPhys.62.1027,Marsiglio2020Jun,Chubukov2020Jun}.

Eqn.~\eqref{AB} makes apparent that the general form of the Josephson critical current is strongly dependent on both the superconducting gap and normal-state conductivity $\sigma_N=1/R_N$ in the junction. If a TLS locally modifies either of these quantities, then the critical current will also be locally modified. To this end, we define the local critical current at site ${\bf r}_i$ of the junction to be $I_c({\bf r}_i)$, which is dependent on the local gap $\Delta({\bf r}_i,\omega)$ and the local density of states $N({\bf r}_i,\,\omega)$. We will reserve discussion of TLSs in the superconducting channel to the next section, and for now focus on the influence of a single TLS in the amorphous oxide tunnel junction. Whereas TLSs in the bulk tunnel barrier form by virtue of the amorphous structure of the oxide~\cite{DuBois2013Feb,Gordon2014Dec,Zeng2016Jul}, additional TLSs in the interface of the substrate and the junction may originate from residual defects~\cite{Holder2013Aug} and the fabrication process itself~\cite{Pop2012Jan,Quintana2014Aug,Zeng2015Apr,Fritz2018May}. Moreover, recent experiments and first principles simulations suggest that electrons at the interface may form a 2D free electron gas~\cite{Jung2009Sep,Zeng2016Jul,Tyner2025}, motivating us to first consider how a single TLS modifies the LDOS via the scattering of normal electrons. 
%For this section, we will then consider the gap to be spatially homogeneous. 
% we will initially consider how a TLS modifies the LDOS in the normal channel. 

The local density of states at the junction interface may be written in terms of the local Green's function $N({\bf r}_i,\,\omega)=-1/(\pi)\Im \{\delta G({\bf r}_{i},{\bf r}_{i}\,\,\omega) \}$, where the full Green's function is given by

\begin{align}
    G({\bf r}_{i},\,{\bf r}_{i};\,\omega)&=G^0({\bf r}_i,\,{\bf r}_i;\,\omega)\notag\\
    &+\sum_{j,\ell}G^{0}({\bf r}_{i},\,{\bf r}_{j};\,\omega)\Sigma({\bf r}_{j},{\bf {\bf r}_{\ell}};\,\omega)G^{0}({\bf r}_{\ell},\,{\bf r}_{i};\,\omega).\label{Green}
\end{align}

\noindent In the above, we 
%define ${\bf r}_{ij}\equiv {\bf r}_i-{\bf r}_j$ and 
have taken $\omega\equiv \omega+i\delta$ with infinitesimal $\delta$ for convenience. The change in the free-electron Green's function from the localized phonon mode is described by the self-energy $\Sigma({\bf r}_j,\,{\bf r}_\ell;\,\omega)$, which may be written (to leading order in the electron-phonon coupling strength $g$) in terms of the bare Green's function $G^0({\bf r}_{j},\,{\bf r}_\ell ;\,\omega)$ and the phonon propagator $D({\bf r}_{j},\,{\bf r}_\ell ;\,\omega-\omega')$ describing the TLS~\cite{AGDBook,MahanBook}:

% \begin{figure}
% %\subfloat[\label{sfig:testa}]{%
%  \hspace{-6mm} \includegraphics[width=1\linewidth]{Interface.pdf}%
% %}
% \caption{\small Change in the local density of states $\delta N(r)/N$ for Al, Ta, and Nb from a single TLS in the amorphous interface located $r=0$. The change in the density of states for Nb is greater than Al and Ta due to a larger electron-phonon coupling strength in Nb.
% }
% \label{}
% \end{figure}

\begin{align}
&\Sigma({\bf r}_{j},\,{\bf r}_\ell \,;\omega)\notag\\
&=ig^2\int \dfrac{d\omega'}{2\pi}\,D({\bf r}_{j},\,{\bf r}_\ell ;\,\omega-\omega')G^{0}\left({\bf r}_{j},\,{\bf r}_\ell ;\,\omega'\right).\label{self}
\end{align}

% \begin{align}
% &\Sigma({\bf r}_{j\ell}\,;i\omega_{n})\notag\\
% &=-\dfrac{g^{2}}{N(0) \beta}\sum_{m}D^{0}({\bf r}_{j\ell};i\omega_{n}-i\omega_{m})G^{0}\left({\bf r}_{j\ell};i(\omega_{n}-\omega_{m})\right),
% \end{align}

\noindent We will model both the phonons in the material of the junction and the TLS itself as Einstein phonons with a constant frequency and simple real-space dependence described by the phonon propagator

\begin{align}
        D({\bf r}_j,\,{\bf r}_\ell;\,\nu)\equiv D(\nu)=\dfrac{2\omega_{0}}{\nu^2-\omega_{0}^{2}},
\end{align}

\noindent where $\nu\equiv \nu+i\delta$ is the bosonic frequency. For the phonons intrinsic to the material, the propagator is assumed to be spatially homogeneous, with an Einstein frequency $\omega_0\equiv\omega_E$ typically in the THz range. For the TLS, the phonon mode is localized on a single lattice site, and has a frequency $\omega_0\equiv\Omega$ around $1$-$10$ GHz~\cite{Muller2019Oct}.

\begin{figure}
%\subfloat[\label{sfig:testa}]{%
 \hspace{-6mm} \includegraphics[width=1\linewidth]{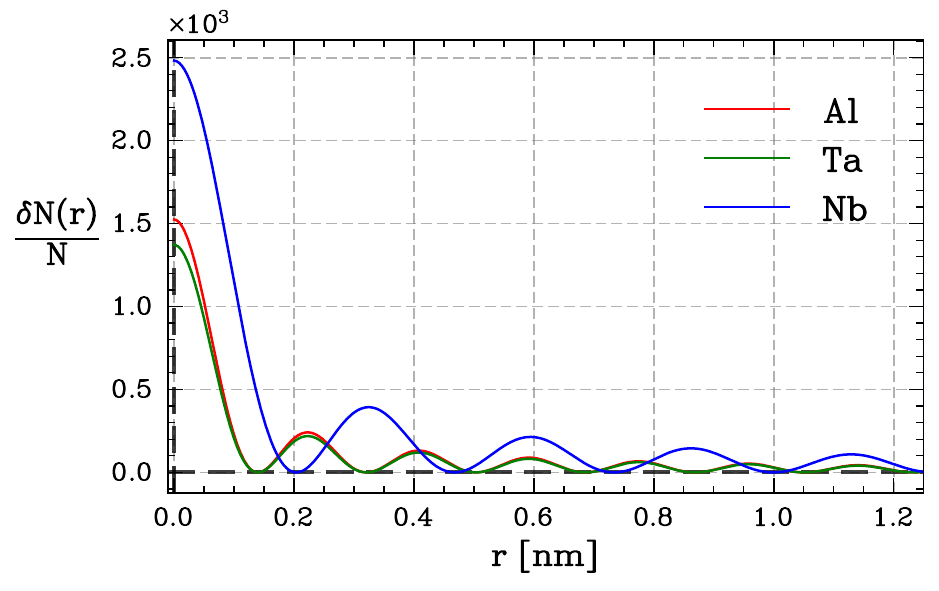}%
%}
\caption{\small Change in the local density of states $\delta N(r)/N$ for Al, Ta, and Nb from a single TLS in the amorphous interface located $r=0$. The change in the density of states for Nb is greater than Al and Ta due to a larger electron-phonon coupling strength in Nb.
}
\label{Fig2}
\end{figure}

From the phonon propagator, we may derive the specific form of the electron-phonon spectral function $\alpha^2 F(\nu)$ and, likewise, the dimensionless electron-phonon (e-p) coupling strength $\lambda$~\cite{RevModPhys.62.1027, Bennemann,Marsiglio2020Jun}:

\begin{align}
    \lambda=2\int_{0}^{\infty}d\nu\dfrac{\alpha^{2}F(\nu)}{\nu}=\dfrac{2N(0)|g|^{2}}{\omega_0}\label{ep_coupling},
\end{align}

\noindent where $N(0)$ is the unperturbed density of states and $\alpha^2F(\nu)=N(0)|g|^{2}\delta(\omega-\omega_0)$ for the Einstein phonon model. For materials such as Al, Ta, and Nb in their crystalline form, the "bare" electron-phonon coupling $\lambda^{(0)}$ is a fixed number, going as $\lambda^{(0)}\sim 0.43$, $0.69$, and $1.11$, respectively~\cite{RevModPhys.62.1027}. However, note that the electron-phonon coupling strength Eqn.~\eqref{ep_coupling} is often found to be inversely proportional to the local phonon frequency~\cite{Lee2006Aug,Balatsky2006Sep,Niestemski2007Dec,Jenkins2009Nov,Fasano2010Oct,Shan2012May}. This suggests that the presence of TLSs in the junction introduces an inhomogeneous landscape of Einstein frequencies and therefore, assuming an anti-correlation between $\omega_0$ and $\lambda$, results in an inhomogeneous landscape of local electron-phonon coupling strengths~\cite{Plakida1987Dec,Balatsky2006Sep,Rosenstein2019Aug}. In general, we expect that the e-p coupling in close proximity to the low-frequency TLS will be amplified relative to the intrinsic $\lambda$ of the material in question. From Eqn.~\eqref{ep_coupling} we see that the local coupling strength at the site of the TLS should go as $\lambda=\lambda^{(0)} \omega_E/\Omega$, where once again we assume that $g$ is roughly homogeneous at the interface and is independent of frequency. This follows from assuming that the boson describing the TLS is not an extended quantized mode~\cite{Balatsky2006Sep}, in addition to the slow variance of $N(0)|g|^2$ typically assumed within the Einstein model~\cite{heid}. We also assume $\lambda^{(0)}$ itself does not vary significantly in the absence of a TLS if we confine ourselves within the superconductor/oxide interface and, later on, within the superconductor. The former assumption is, once again, in agreement with recent numerics predicting the breakdown of amorphous structure at the oxide interface and for materials such as Ta with larger atomic mass~\cite{Wang2024May,Tyner2025}. Finally, we will assume the electric dipole moment associated with the TLS (as discussed in Ref.~\cite{Tyner2025}) does not appreciably modify $\lambda$.
%Physically, a roughly constant value of $\lambda^{(0)}$ and
%~\cite{Balatsky2006Sep}\footnote{Note that, by taking $g$ to be independent of $\Omega_0$, we are assuming that the boson describing the TLS is not an extended quantized mode. In this way, the amplification of $\lambda$ from a single localized mode is distinct from the amplification of $\alpha^2F(\omega)$ in disordered superconductors~\cite{Bergmann1971Jun,Jackle1980Mar}.}. 
From these assumptions, a local amplification of the electron-phonon coupling on the order of two orders of magnitude is expected for common materials of interest like Al, Ta, and Nb. An amplification of the local e-p coupling of comparable magnitude has been observed experimentally in crystalline Au embedded with Ag nanoparticles~\cite{Kumbhakar2024May}. In such a case, large $\lambda$ values have also been attributed to the coupling of electrons to localized phonon modes. For our study, we re-iterate that these local pockets of huge $\lambda$ are a natural consequence of localized, low-frequency TLSs. 

\begin{figure*} 
\subfloat[\label{sfig:testa}]{%
  \includegraphics[width=.42\linewidth]{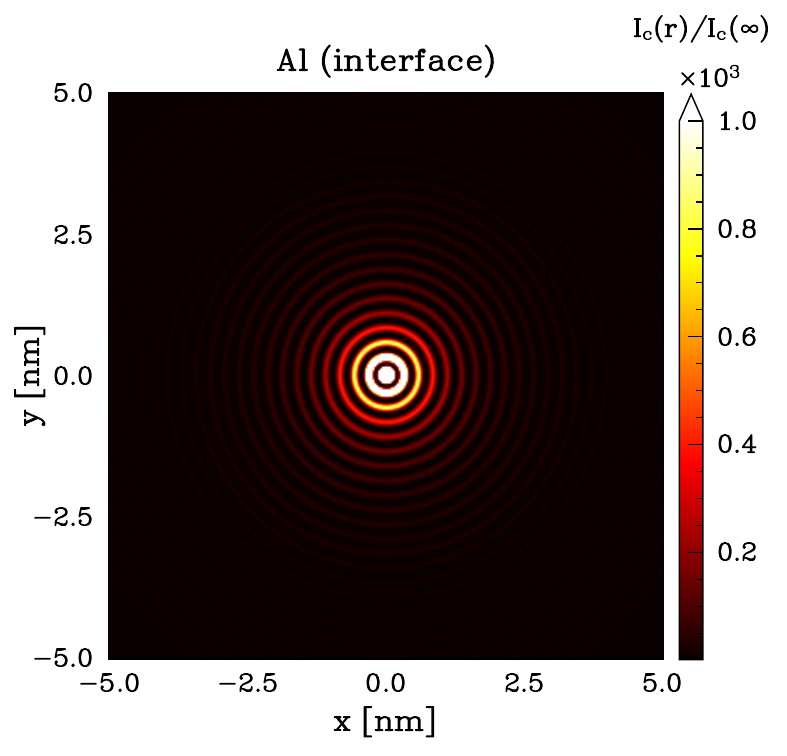}%
}\hspace{10mm}
\subfloat[\label{sfig:testa}]{%
  \includegraphics[width=.42\linewidth]{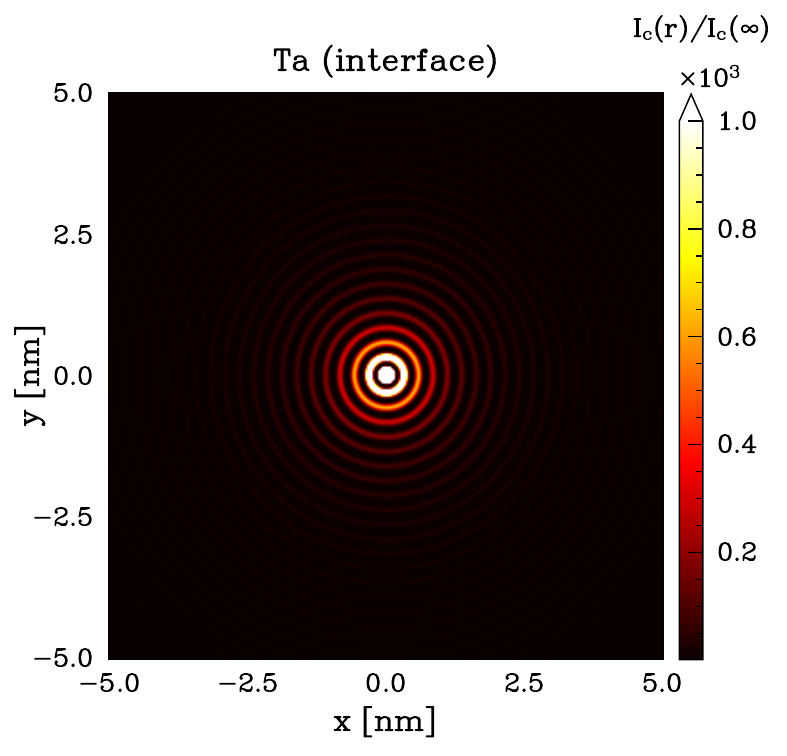}%
}

\subfloat[\label{sfig:testa}]{%
  \includegraphics[width=.42\linewidth]{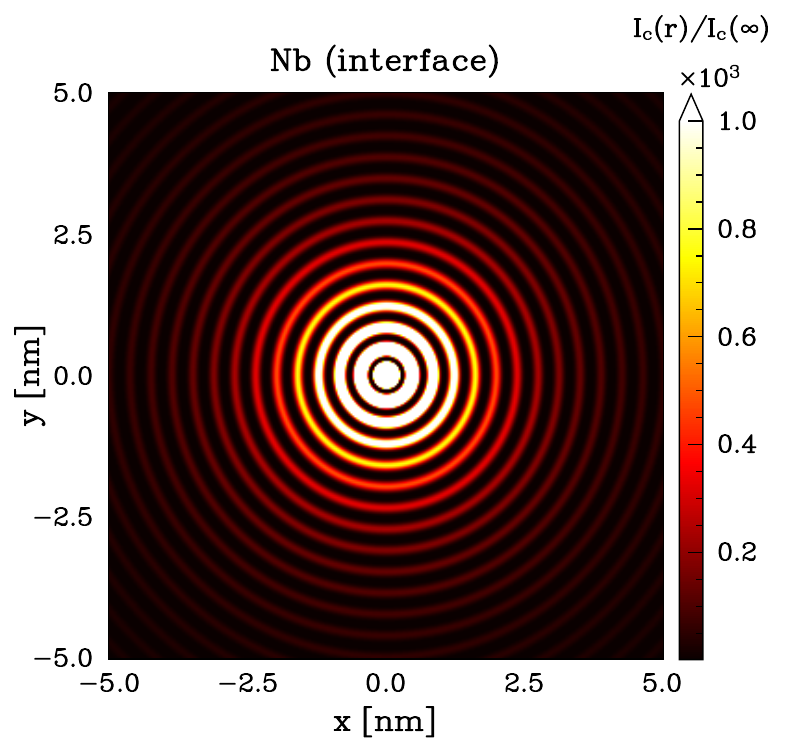}%
}
\caption{Amplification of the local Josephson current $I_c({\bf r})$ at the superconductor/oxide interface for Al (top left), Ta (top right), and Nb (bottom). Real-space picture taken over a 10 nm $\times$ 10 nm square. Over this area, the local Josephson current is amplified by around $4\times 10^1$ for Al and Ta, while for Nb $I_c({\bf r})$ is amplified by around $3\times 10^2$. The oscillatory behaviour is a consequence of the spatial dependence of the real-space Green's function.
}
\label{Fig3}
\end{figure*}

The shift of the local density of states at site ${\bf r}_i$ in the amorphous oxide surface resulting from inelastic scattering of normal electrons off a TLS at site ${\bf r}_\ell$ may be calculated from the above self energy, and is found to be

\begin{align}
\dfrac{\delta N({\bf r}_{i},\omega)}{N(0)}=\dfrac{\pi^{2}}{2}\Omega N(0)\lambda_\ell\mathcal{P}^{2}({\bf r}_{i\ell})\Theta\left(\omega-\Omega_\ell\right)\label{LDOS},
\end{align}
% \begin{align}
% \dfrac{\delta N({\bf r}_{i},\omega)}{N(0)}=\dfrac{\pi^{2}}{2}\Omega_{0}N(0)\sum_{j}\delta_{j\ell}\lambda_j\mathcal{P}^{2}({\bf r}_{ij})\Theta\left(|\overline{\omega}|-1\right)\label{LDOS},
% \end{align}

\noindent where we define ${\bf r}_{ij}\equiv {\bf r}_i-{\bf r}_j$ and $\lambda_\ell$ ($\Omega_\ell$) to be the local electron-phonon coupling (local TLS frequency) at site ${\bf r}_\ell$. For simplicity and consistency, we will drop the $\ell$ subscript on both the e-p coupling and the TLS frequency, and assume all TLS frequencies are identical. In the presence of several TLSs, the change in the LDOS is the same as in Eqn.~\eqref{LDOS}, except now we sum over the TLS sites. In addition, we define

\begin{align}
    \mathcal{P}({\bf r}_{ij})=\begin{cases}
        &J_0(k_F  r_{ij})e^{-r_{ij}/(2l)}\qquad \textrm{2D}\\
        &\\
        & \dfrac{\sin(k_F r_{ij})}{k_F r_{ij}}e^{-r_{ij}/(2l)}\qquad \textrm{3D}\label{modulation}
    \end{cases}
\end{align}

\noindent to be a spatial modulation parameter from the form of the real-space Green's function, with $l$ taken to be the mean-free path of electrons at the interface and $J_0(...)$ the zeroth order Bessel function of the first kind. For the time being, we will take the former form of $\mathcal{P}({\bf r}_{ij})$, as we are considering TLSs at the 2D amorphous/superconductor interface. Note that the LDOS only experiences a shift when $\omega>\Omega$; i.e., when an inelastic scattering channel is opened between normal electrons and the localized mode~\cite{Stipe1998Jun,Balatsky2003Dec,Zhu2004Jul,Balatsky2006May,Behm2010Dec}.
% the frequency crosses a threshold set by the TLS frequency

% The Heaviside step function in Eqn.~\eqref{LDOS} ensures that there is only a finite change to the LDOS for frequencies $\omega>\Omega_0$, as the frequency must cross this threshold in order to open an i

% This is reminiscent of what is seen in IETS experiments, where the increase of a bias voltage beyond a threshold energy excites vibrational modes of localized molecules, thereby opening an inelastic tunneling channel and an additional contribution to the tunneling current~\cite{Stipe1998Jun,Balatsky2003Dec,Zhu2004Jul,Balatsky2006May,Behm2010Dec}. 

In Fig.~\ref{Fig2}, the change in the LDOS from a single TLS is plotted against the distance to the TLS. For each material (Al, Ta, and Nb), the corresponding bare electron-phonon coupling strength $\lambda^{(0)}$ and Einstein frequency $\omega_E$ is taken into account. The full frequency-dependent gap function $\Delta(\omega)$ is found self-consistently via the Eliashberg equations on the imaginary frequency axis~\cite{Marsiglio2018Jul} before performing an exact analytical continuation to the real frequency axis~\cite{Marsiglio1988Apr}. Numerical parameters for the Eliashberg calculation are dependent on $\lambda^{(0)}$ and are determined in such a way to ensure convergence of the Matsubara sum (see Appendix C and Ref.~\cite{Heath2024Jul} for more details). To calculate the LDOS, we uniformly take $\Omega=10$ GHz and assume $N(0)$ to be that of a free electron gas. While there is some small sense of variability of $N(0)$ and $\Omega$, we uniformly take $\Omega N(0)\approx 1$ based upon these assumptions. We see that amplification is maximized in very close proximity to the TLS, scaling as $\omega_E/\Omega$ and dying off as we move away from the localized mode. The amplification for Nb is much larger than Al or Ta, due to the relatively large electron-phonon coupling in the former material. 

Note that the change of the LDOS from a TLS in the bulk amorphous oxide will have a different form from Eqn.~\eqref{LDOS}, which defines the change in the LDOS from a TLS at the interface. For clarity, we define the bulk region of the Josephson junction as the part of the amorphous oxide that is significantly farther from the 2D interface than the depth of penetration into the barrier. In this regime, the oxide is characterized by a large electronic energy gap $\epsilon_g$, and as such Eqn.~\eqref{LDOS} is modified by an additional Heaviside step function $\Theta\left({\omega}-\epsilon_g\right)$ which takes into account the gap of the amorphous material. As the band gap of amorphous oxides is on the order of $10^3$ THz, the contribution to the local Josephson current from a TLS in the amorphous oxide bulk is negligible compared to that of a TLS at the superconductor/oxide interface. This follows physically simply because, within the insulating oxide, there are no normal electrons with energy below $\epsilon_g$ to scatter off the low-frequency TLS.

% \begin{align}
% \dfrac{\delta N({\bf r}_{i},\omega)}{N(0)}=\dfrac{\pi^{2}}{2}\Omega_{0}N(0)\sum_{j}\lambda_j\mathcal{P}^{2}({\bf r}_{ij})\Theta\left(|\overline{\omega}|-\overline{\epsilon_g}\right).
% \end{align}
% \begin{figure*} 
% \subfloat[\label{sfig:testa}]{%
%   \includegraphics[width=.42\linewidth]{Pic3A.pdf}%
% }\hspace{10mm}
% \subfloat[\label{sfig:testa}]{%
%   \includegraphics[width=.42\linewidth]{Pic3B.pdf}%
% }

% \subfloat[\label{sfig:testa}]{%
%   \includegraphics[width=.42\linewidth]{Pic3C.pdf}%
% }
% \caption{Amplification of the local Josephson current at the superconductor/oxide interface for Al (top left), Ta (top right), and Nb (bottom). Real-space picture taken over a 10 nm $\times$ 10 nm square. Over this area, the local Josephson current is amplified by around $4\times 10^1$ for both Al and Ta, while for Nb the Josephson current is amplified by around $3\times 10^2$. The oscillatory behavior is a consequence of the spatial dependence of the real-space Green's function.}
% \label{Fig3}
% \end{figure*}

The total local conductivity in the normal state can be written as $\sigma_N({\bf r}_i,\omega)=4\pi N_1({\bf r}_i,\,\omega)N_2({\bf r}_i,\,\omega)\langle |T|^2\rangle$,
where $\langle |T|^2\rangle$ is the averaged elements of the transition matrix from the left to the right of the oxide and $N_\alpha({\bf r}_i,\,\omega)=N_\alpha(0)+\delta N_\alpha({\bf r}_i,\,\omega)$, where $\alpha=1,2$ denotes the electrons from either side of the junction barrier~\cite{Kulik2}\footnote{Due to the relative thinness of the amorphous oxide bulk~\cite{Rippard2002Jan,Yang2020Sep,Connolly2024May}, the influence of normal quasiparticles tunnelling through the oxide near high transmission regions (such as defects)~\cite{Kurter2022Mar}, and the re-emergence of crystalline order in materials such as Ta~\cite{Wang2024May,Tyner2025}, we assume that there is a finite overlap of $N_1({\bf r}_i,\,\omega)$ and $N_2({\bf r}_i,\,\omega)$ in close proximity to the oxide interface. Therefore, a TLS on or near the interface will influence the LDOS of both electron populations.}. The resultant change in $I_c({\bf r}_i)$ is then

\begin{figure}[t!]
%\subfloat[\label{sfig:testa}]{%
 \hspace{-6mm} \includegraphics[width=1\linewidth]{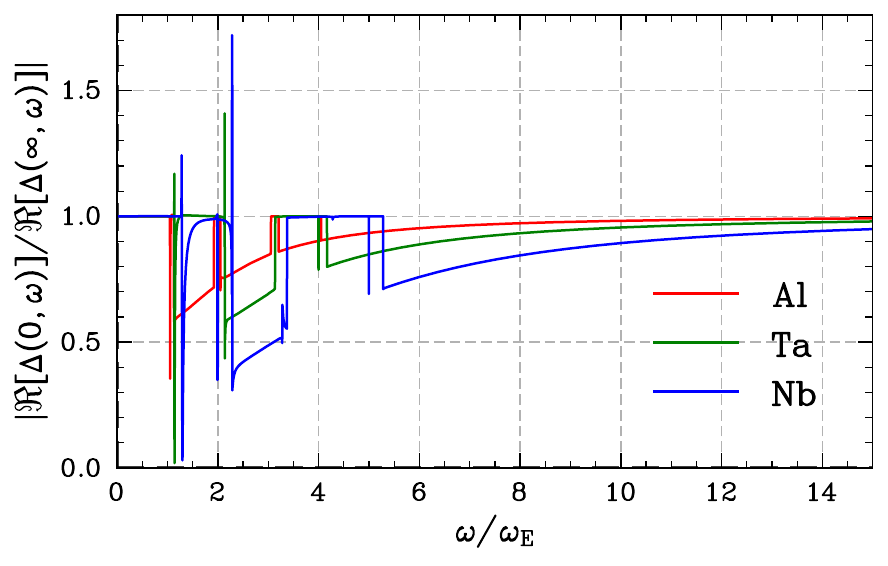}%
%}
\caption{\small Change of the local real gap function $\Delta(0,\,\omega)$ in close proximity to the TLS. 
The net effect of a TLS in the superconducting bulk results in a decrease in $\Delta(0,\,\omega)$ across the frequency range. 
}
\label{Fig4}
\end{figure}
\phantom{dd dfj;kdfj  ;jfsjf }
%\begin{widetext}
\begin{align}
    \delta I_c({\bf r}_{i})&=\dfrac{2}{\pi e R_N} \int_{-\infty}^\infty d\omega \dfrac{\delta N({\bf r}_{i},\,\omega)}{N(0)}\left(1+\dfrac{1}{2}\dfrac{\delta N({\bf r}_{i},\,\omega)}{N(0)}\right)\notag\\
    &\phantom{=\dfrac{2}{\pi e R_N}\int_{-\infty}^\infty d\omega}\times\bigg[\Re \left\{ \dfrac{\Delta(\omega)}{\sqrt{\omega^2-\Delta^2(\omega)}}
    \right\}\bigg]^2,
\end{align}
\phantom{dd dfj;kdfj  ;jfsjf df dsf fdsf sd dfd dfsdf sdf dsf where}
%\end{widetext}
\noindent where we now drop the subscript $\alpha$.
%differentiating the LDOS of electrons from the left or right of the oxide. 
%In the above, note that the spatial dependence of the gap is not considered, as we are (for now) purely considering a TLS in the normal channel. 
%However, the full frequency dependence of the gap is imperative to include, as now the LDOS carries frequency dependence from the presence of the TLS. 
We write the change of the critical current in the unitless form $I_c({\bf r}_i)/I_c(\infty)=1+\delta I_c({\bf r}_i)/I_c(\infty)$, where $I_c(\infty)$ is the critical current infinitely far away from the TLS (i.e., Eqn.~\eqref{AB}). The change in $I_c({\bf r}_i)$ is shown in Fig.~\ref{Fig3} as a real-space density plot for Al, Ta, and Nb. Whereas the results for Al and Ta are appreciable to each other, the amplification of $I_c({\bf r}_i)$ in Nb is around an order of magnitude greater than the former two materials. This suggests that the local modulation of the critical current from a single TLS is amplified for materials with a larger bare electron-phonon coupling strength $\lambda^{(0)}$. 

% This is consistent with the fact that Nb has a larger electron-phonon coupling strength then Al and Ta.
%This is a direct consequence of Nb having an electron-phonon coupling strength noticeably larger than that of Al and Ta. 
%The results of this section suggest that individual TLSs may be detected experimentally through IETS. An STM tip may be held constant over the oxide surface and perform a sweep of the bias voltage, from which the amplified LDOS can be read off from the first derivative of the I-V curve. Such an amplification should dwarf other sources of thermal and quantum fluctuations that smear the I-V characteristics of the Josephson junction~\cite{Martinis1987Apr,Martinis1988Aug,Krasnov2007Dec}. The amplification of the Josephson critical current itself may also be determined by the I-V data~\cite{Kautz1990Dec}. The more imperative issue of qubit decoherence from the critical current amplification will be addressed in Section IV. 
%Similarly, MENTION HERE LARGER JUNCTION
%\newpage

%\phantom{dfafd}
\section{III. Two-level systems in the superconductor}

In the previous section, we considered the local modulation of the critical Josephson current from a single TLS in the amorphous oxide/superconductor interface. Now, we will consider the effect of a TLS in the superconductor itself~\cite{Grunhaupt2019Aug,Kristen2024May,Yang2020Sep}. Assuming a single TLS is located in the superconducting bulk, the change in the LDOS is the same as before, as this is calculated in the normal state. However, the TLS will influence the gap if it has a finite magnetic dipole moment. %Many TLSs are known to be characterized by a finite electrical dipole moment~\cite{DuBois2013Feb,Holder2013Aug,Gordon2014Dec} in addition, in some cases, to a magnetic dipole moment~\cite{Bialczak2007Nov,Sheridan2021Nov,Huang2024Dec}. 
It has been noted that a TLS in the superconducting bulk would most likely be paramagnetic or be induced by a nearby magnetic impurity~\cite{Bialczak2007Nov,Yang2020Sep,deGraaf2020Dec,Sheridan2021Nov,Kristen2024May,Huang2024Dec}, therefore behaving like a paramagnetic impurity and subsequently suppressing the local superconducting gap function~\cite{Anderson1959Sep,Parks2,Balatsky2006May}. Indeed, fluctuations of the gap observed in disordered NbN and similar materials~\cite{Lemarie2013May,deGraaf2017Jan,Saveskul2019Nov,Liao2019Dec,Carbillet2020Jul} suggest the ubiquitous presence of paramagnetic impurities in superconducting samples. As such, we will consider the effect of a TLS in the superconductor as if it were a local paramagnetic impurity which directly influences the superconducting gap function. 
%Note that magnetic moments near the oxide layer are severely suppressed for Ta$_2$O$_{5-x}$ as compared to Nb$_2$O$_{5-x}$~\cite{Pritchard2024Oct}, leading to the possibility that the influence of the TLS in the superconducting channel is suppressed for qubits made from Ta.

We write the local Josephson critical current as $I_c({\bf r}_i)={I_c}^{(0)}({\bf r}_i)+\delta I_c({\bf r}_i)$, where
\begin{widetext}
\begin{subequations}
\begin{equation}
    I_c({\bf r}_{i})=\dfrac{2}{\pi e R_N} \int_{-\infty}^\infty d\omega \Re \left\{ \dfrac{\Delta(\omega)}{\sqrt{\omega^2-\Delta^2(\omega)}}
    \right\}\Re \left\{ \dfrac{\Delta({\bf r}_i,\,\omega)}{\sqrt{\omega^2-\Delta^2({\bf r}_i,\,\omega)}}
    \right\},\label{SCI1}
\end{equation}
\begin{equation}
    \delta I_c({\bf r}_{i})=\dfrac{2}{\pi e R_N} \int_{-\infty}^\infty d\omega \dfrac{\delta N({\bf r}_{i},\,\omega)}{N(0)}\left(1+\dfrac{1}{2}\dfrac{\delta N({\bf r}_{i},\,\omega)}{N(0)}\right)\Re \left\{ \dfrac{\Delta(\omega)}{\sqrt{\omega^2-\Delta^2(\omega)}}
    \right\}\Re \left\{ \dfrac{\Delta({\bf r}_i,\,\omega)}{\sqrt{\omega^2-\Delta^2({\bf r}_i,\,\omega)}}
    \right\}\label{SCI2}.
\end{equation}
\end{subequations}
\end{widetext}

\begin{figure*} 
\subfloat[\label{sfig:testa}]{%
  \includegraphics[width=.42\linewidth]{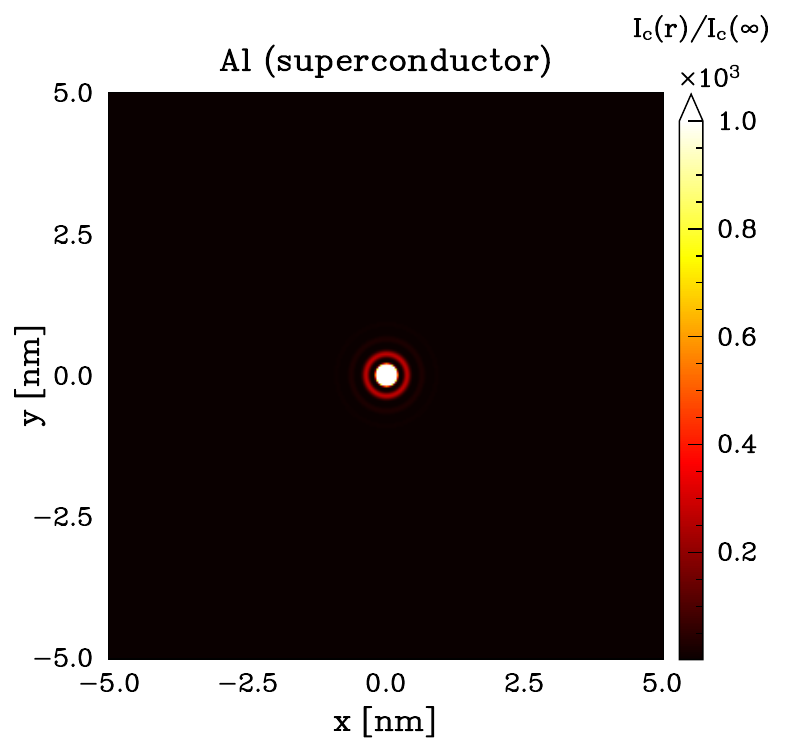}%
}\hspace{10mm}
\subfloat[\label{sfig:testa}]{%
  \includegraphics[width=.42\linewidth]{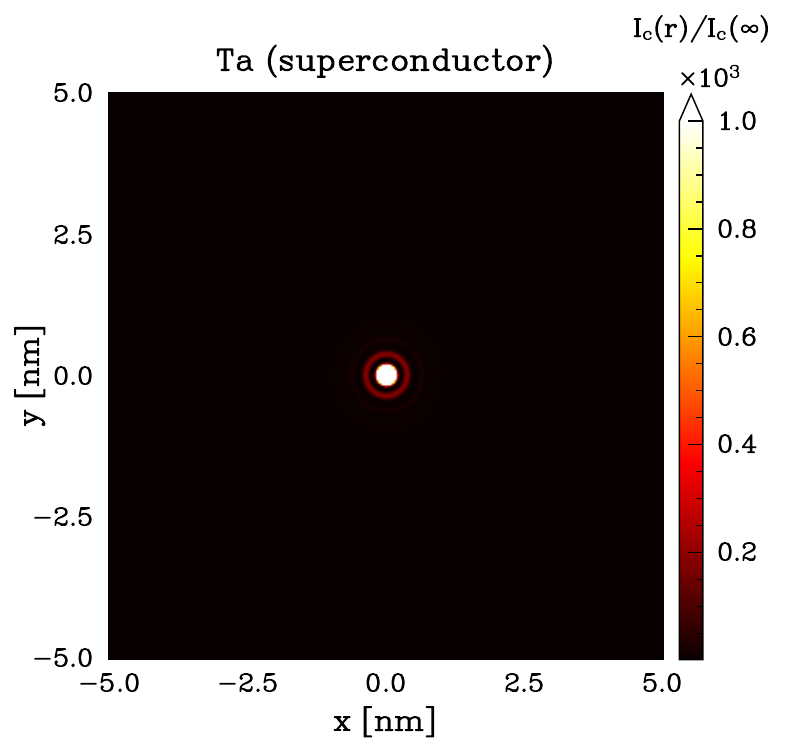}%
}

\subfloat[\label{sfig:testa}]{%
  \includegraphics[width=.42\linewidth]{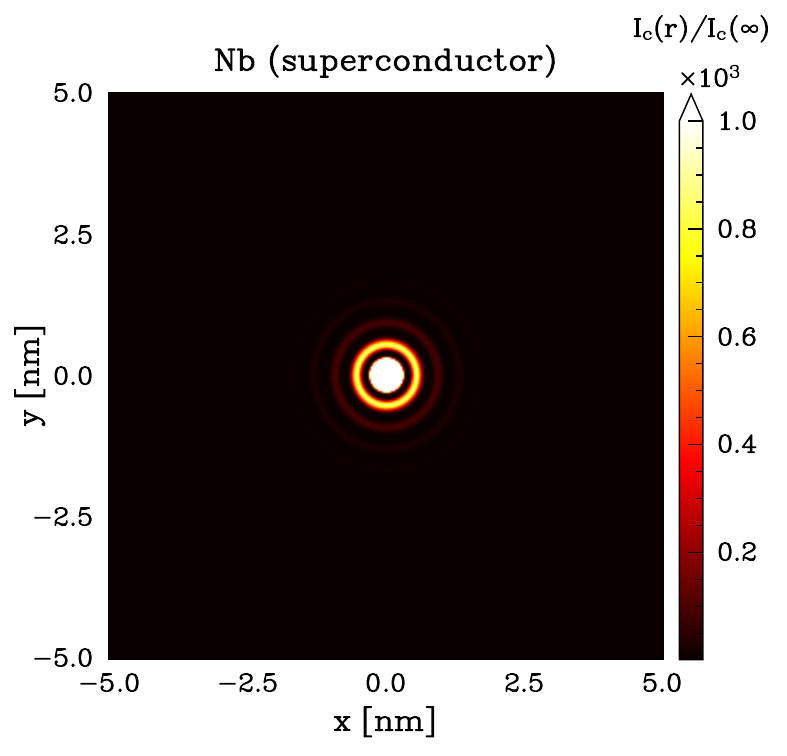}%
}
\caption{Amplification of the local Josephson current in the superconducting bulk for Al (top left), Ta (top right), and Nb (bottom). Real-space picture taken over a 10 nm $\times$ 10 nm square. Over this area, the Josephson critical current is amplified by a factor of around $10^1$ for Al and Ta, while for Nb the critical current is amplified by $1.3\times 10^2$.}
\label{Fig5}
\end{figure*}

\noindent In the above, $\delta I_c({\bf r}_i)$ denotes the change purely from the modification of the LDOS. Both terms incorporate the change of the local gap function. In the case of a TLS at the amorphous interface, note that ${I_c}^{(0)}({\bf r}_i)={I_c}^{(0)}(\infty)$, as the gap function will remain spatially homogeneous. 
%Also note that, if assume that the TLS is in close proximity to the amorphous interface, a suppression of the gap could result in high transmission paths for normal quasiparticles to tunnel through the junction~\cite{Kurter2022Mar}.
%the TLS, while in the superconductor, is in close proximity to the amorphous interface, and thus the amplified LDOS in the normal phase will result in a high transmission region of the junction for normal particles, thereby influencing the LDOS on both sides of the amorphous junction. This is especially relevant for a TLS in the superconducting channel, as a suppression of the gap will result in high transmission paths for normal quasiparticles to tunnel through the junction~\cite{Kurter2022Mar}.

In Eqns.~\eqref{SCI1} and \eqref{SCI2}, the local behaviour of the gap function and the related renormalisation function $Z({\bf r}_i,\,\omega)$ is given by

\begin{subequations}
\begin{equation}
\dfrac{\Delta({\bf r}_i,\,\omega)}{\Delta(\infty,\,\omega)}=\dfrac{Z(\infty,\,\omega)}{Z({\bf r}_i,\,\omega)}-\mathcal{P}({\bf r}_i)\dfrac{\omega\Im[Z(\infty,\,\omega)]}{Z({\bf r}_i,\,\omega)\sqrt{\Delta^{2}(\infty,\,\omega)-\omega^{2}}},\label{Gap_imp}
\end{equation}
    \begin{equation}
\dfrac{Z({\bf r}_i,\,\omega)}{Z(\infty,\,\omega)}=1+\mathcal{P}({\bf r}_i)\dfrac{\omega\Im[Z(\infty,\,\omega)]}{Z(\infty,\,\omega)\sqrt{\Delta^{2}(\infty,\,\omega)-\omega^{2}}},\label{Z_imp}
    \end{equation}
\end{subequations}

\noindent where $\mathcal{P}({\bf r}_i)$ is given in Eqn.~\eqref{modulation} and we have used the inelastic scattering rate $\Gamma({\bf r}_i,\,\omega)=\mathcal{P}({\bf r}_i)\omega \Im [Z({\bf r}_i,\,\omega)]\approx \mathcal{P}({\bf r}_i)\omega \Im [Z(\infty,\,\omega)]$~\cite{Marsiglio1992May,Balatsky2006May,Heath2024Jul}. In the above, the renormalisation function $Z({\bf r}_i,\,\omega)$ is related to the odd component of the normal self energy. The functional dependence on ${\bf r}_i$ given in Eqn.~\eqref{Z_imp} is the same with or without a finite magnetic dipole moment. In the absence of a magnetic dipole moment, Eqn.~\eqref{Gap_imp} reduces to $\Delta(r,\,\omega)=\Delta(\infty,\,\omega)\equiv \Delta(\omega)$~\cite{Anderson1959Sep,Parks2}. We will assume that the gap function at ${\bf r}_i=0$, while amplified from the low-frequency TLS, is severely suppressed due to the on-site impurity~\cite{Zhu2016}.

% , and as a consequence the local change in the Josephson current is identical to what was previously found in the normal channel. As mentioned previously, however, TLSs in the superconducting channel will most likely be paramagnetic or be induced by a nearby magnetic impurity, and thus the gap will more than likely be modified from the presence of a TLS in the superconductor itself.

Results for the change in the local Josephson current from a TLS in the superconducting channel is shown in Fig.~\ref{Fig5} where, once again, we assume the TLS is associated with some finite magnetic dipole. As in the previous section, the local structure of the gap and renormalisation functions is determined self-consistently from the Eliashberg equations on the imaginary frequency axis before analytically continuing to the real frequency axis. In Fig.~\ref{Fig5}, note that the amplification of $I_c({\bf r}_i)$ is now suppressed compared to what we previously saw in the normal channel, as a direct result of the suppression of $\Delta({\bf r}_i,\,\omega)$ (see Fig.~\ref{Fig4}). The range of the critical current amplification from a TLS in the superconductor is also much smaller than that of a TLS at the superconductor/oxide interface. This suggests that signatures of TLSs in the superconductor would be harder to detect than TLSs in the normal channel. The effects of a TLS in superconducting Nb, however, should still have some appreciable effect on the local Josephson current environment. Similarly, note that magnetic moments near the oxide layer are severely suppressed for Ta$_2$O$_{5-x}$ as compared to Nb$_2$O$_{5-x}$~\cite{Pritchard2024Oct}, leading to the possibility that the influence of a TLS in the superconducting channel is suppressed for qubits made from Ta.

\section{IV. Decoherence of a phase qubit from a locally amplified critical current}

We now turn our attention to how an amplified Josephson current may induce decoherence in a superconducting qubit. While it is well known that TLSs are parasitic to high-quality qubits, typical decoherence channels may be rectified through a detailed characterization of the individual TLS-qubit coupling~\cite{Mansikkamaki2024Apr}. Here, we illustrate how many-body effects of low-frequency TLSs can lead to qubit decoherence, {\it independent} of the specific qubit or TLS frequency.

For simplicity, we will take the example of a phase qubit. The phase qubit is a simple and well-known example of a superconducting qubit, with a Hamiltonian given by

\begin{align}
    H=\dfrac{Q^{2}}{2C}-\dfrac{I_{c}\Phi_{0}}{2\pi}\cos\delta-\dfrac{I_{b}\Phi_{0}}{2\pi}\delta,\label{PhaseHam}
\end{align}

\noindent where $Q$ is the charge on the junction plates with capacitance $C$. As $\delta$ and $Q$ are conjugate variables, we can quantize the above Hamiltonian and expand the potential about one of the minima, obtaining a qubit with a Bloch vector controlled by the bias current $I_b$. In the "resistively- and capacitively-shunted junction” (RCSJ) picture, the Josephson junction describes a phase particle of effective mass $C\cdot (\Phi_0/2\pi)^2$ moving in a potential $U(\delta)=\epsilon_J(-I_b/I_0-\cos \delta)$. This "washboard potential" is characterized by periodic minima with depth $\Delta U\sim \epsilon_J (1-I_b/I_c)^{3/2}$. The oscillation of the phase about one of the minima of $U(\delta)$ defines the lowest resonant frequency of the qubit $\omega_{01}=\sqrt{{2\pi I_{c}}/{(\Phi_{0}C)}}\left[2\left(1-{I_{b}}/{I_{c}}\right)\right]^{1/4}$. Note that the range of the bias current $I_b$ for the phase qubit is very limited~\cite{Martinis2002Aug}. If the bias current is too small compared to the critical current $I_c$, then the depth of the local potential is increased, leading to smaller energy level spacings and making it harder to experimentally distinguish the ground state from higher energy levels. If the bias current is too large, then the local potential is too shallow, leading to tunnelling of the phase particle out of the well. As such, an experimentally relevant ratio of $I_b$ to $I_c$ may be approximated as $I_b/I_c\approx 0.988$ for an energy level splitting $\omega_{01}\approx 10$ GHz and critical current $I_c\approx 10$ $\mu$A.

\begin{figure}
%\subfloat[\label{sfig:testa}]{%
 \hspace{0mm} \includegraphics[width=1\linewidth]{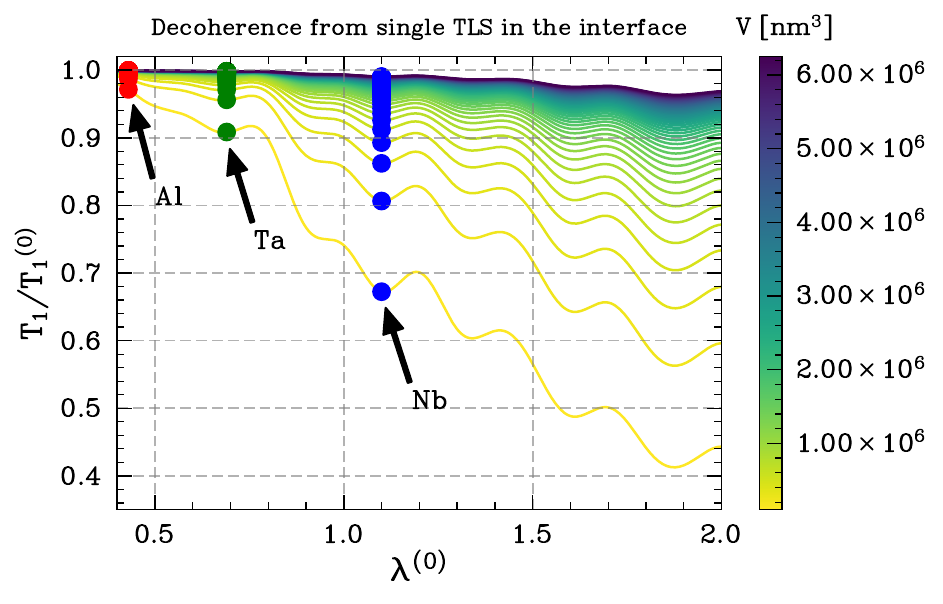}\\
\includegraphics[width=1\linewidth]{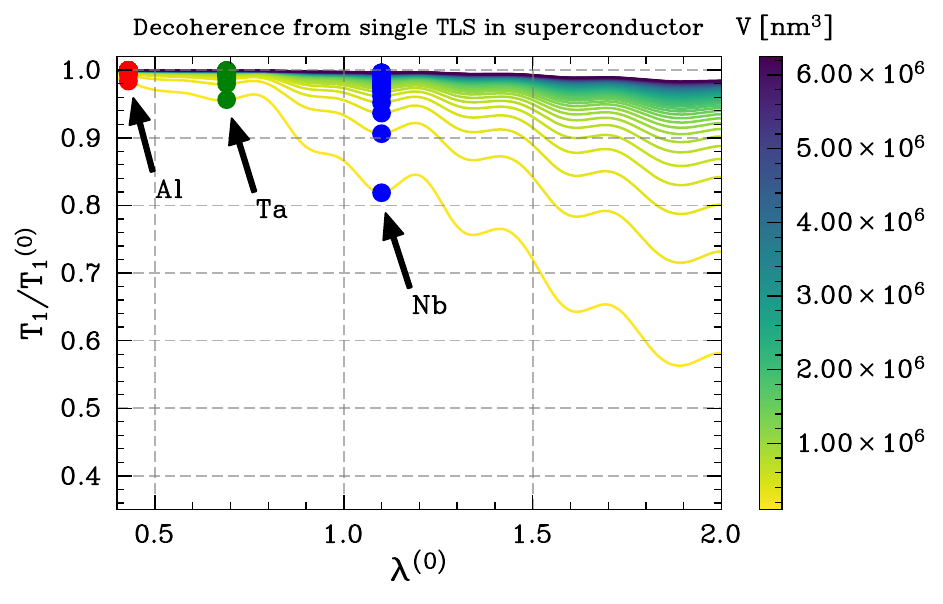}%
%}
\caption{\small The change in the $T_1$ time from a single TLS in the interface (top) and in the superconductor (bottom) plotted versus the bare electron-phonon coupling strength $\lambda^{(0)}$ of the junction material. Data plotted for several different sizes of the superconductor.}
\label{Fig6}
\end{figure}

Fluctuations in the bias current from the external environment induces longitudinal relaxation of the Bloch vector, which may be quantified by the $T_1$ relaxation time:

\begin{align}
    {1}/{T_1}=\omega_{01}\alpha\tan\delta.\label{T1_eqn}
\end{align}

\noindent Here, $\alpha$ is the ratio of the real capacitance to the total capacitance of the dielectric in the junction and $\tan\delta$ is the loss tangent of the junction dielectric (i.e., the ratio of the imaginary to real permittivities). From Eqn.~\eqref{T1_eqn}, one can see that the $T_1$ time is sensitive to the energy-level splitting $\omega_{01}$ of the qubit and, as such, the critical current $I_c$. Any amplification of the average $I_c$ (say, from a TLS) will result in a deeper local minimum of the washboard potential. As a consequence, the lowest resonant frequency of the qubit will become indistinguishable from higher resonant frequencies, as the dynamics of the system will no longer be restricted to the Hilbert space spanned by the two lowest energy states~\cite{Martinis2003Mar}. Any noise in the bias current will then be more likely to induce energy transitions to higher frequencies.

Following the discussion from the previous two sections, we'll take the $T_1$ time before and after the introduction of a TLS to be $T_1^{(0)}$ and $T_1$, respectively. The ratio of the latter to the former is then

\begin{align}
    \dfrac{T_1}{T_1^{(0)}}&=\sqrt{\dfrac{I_{c}^{(0)}}{I_{c}}}\dfrac{\left(1-{I_{b}}/{I_{c}^{(0)}}\right)^{1/4}}{\left(1-{I_{b}}/{I_{c}}\right)^{1/4}}\approx \sqrt{\dfrac{I_{c}^{(0)}}{I_{c}}},
\end{align}
where we have assumed $I_b/I_c^{(0)}\approx I_b/I_c\lessapprox 1$; that is, the bias current is tuned to the optimal value in the qubit with and without the TLS. Results for the change in the $T_1$ time from a single TLS in the oxide/superconductor interface and in the superconductor itself is shown in Fig.~\ref{Fig6} as a function of the electron-phonon coupling strength. Note that we uniformly take $\omega_E=5$~THz, $k_F=20$~nm$^{-1}$, and $\ell=15$~nm for all results reported in Fig.~\ref{Fig6}. We also consider several different sizes of the superconducting material itself, from $10\times 50\times 50$ nm$^3$ to $50\times 250\times 250$ nm$^3$. 

For both scenarios of TLSs in the interface and in the superconducting bulk, we see that the $T_1$ time is reduced as the e-p coupling increases. For a TLS in the interface, Al and Ta appear to be affected for smaller junctions, with $T_1$ reduced by up to $90$\% its original value. Decoherence from a single TLS is more severe for Nb, and becomes considerable for larger junctions as $\lambda$ increases. For a TLS in the superconductor, the overall effect on Al and Ta is negligible, even for smaller junctions. However, the overall effect on Nb and other materials with larger electron-phonon coupling strength is still appreciable for smaller junctions.  

% \newpage \phantom{.}
% \newpage 

\section{V. Conclusions}

In this work, we have investigated the effect a single TLS will have on the local Josephson critical current. We have done this by building off of recent {\it ab initio} simulations which suggest that soft phonon modes are a likely candidate for TLSs~\cite{Tyner2025}. For a TLS in the amorphous oxide/superconductor interface, the presence of the TLS results in a new inelastic scattering channel between normal electrons and the low-frequency mode, which in turn results in a short-range amplification of the local density of states. This increase in the LDOS results in an amplified local Josephson critical current, with a magnitude and real-space range dependent on the characteristic phonon frequency and electron-phonon coupling strength of the junction material. We find that this local amplification of the critical current is larger for Nb compared to Al and Ta, and in general will become more exaggerated if we consider materials with larger electron-phonon coupling. We then consider a single TLS in the superconductor itself, and assume that the TLS is associated with a magnetic dipole moment which reduces the local superconducting gap. While the amplification of the critical current from the TLS still persists in this case, it is negligible for Al and Ta and is characterized by a shorter range compared to what we find for a TLS at the interface. 

Taking the example of a phase qubit, we find that a local amplification of the critical current has an adverse effect on the dephasing $T_1$ time. Specifically, we find that a single TLS is enough to open a new decoherence channel in the phase qubit that is independent of resonance-based population decay~\cite{Ku2005Jul}. Such decoherence becomes amplified for smaller junctions and for junctions manufactured from compounds with larger electron-phonon coupling strength. In such a limit, a new decoherence channel opens from TLSs in the superconducting bulk. To the best of our knowledge, this is the first identification of a pronounced decoherence mechanism unique to the superconducting medium of the qubit, and will be a significant source of decoherence for smaller junctions. Also note that our study is restricted to the case of a {\it single} TLS. Recent electric field and swap spectroscopy measurements~\cite{Cooper2004Oct,Lisenfeld2015Feb,Bilmes2021Feb} suggest a finite density of TLSs in contemporary superconducting qubits in addition to a dominant loss channel originating from surface defects~\cite{Lisenfeld2019Nov,Bilmes2021Aug}. As a consequence, our present study sets a lower estimate for qubit decoherence from inelastic scattering of electrons off TLSs.

%We identify a unique, materials-dependent decoherence mechanism that will affect superconducting qubits regardless of the specific qubit frequency. 
Mitigation of this decoherence channel would involve a careful choice of material and device parameters for the Josephson junction. Our results suggest that junctions constructed from materials with a smaller intrinsic electron-phonon coupling strength (such as Al or Ta) will not be as affected as those of materials with a larger coupling (such as Nb). This is a potential problem for current superconducting qubit design, where materials with a larger electron-phonon coupling (and, hence, a larger critical temperature and superconducting gap) are often sought after due to negligible decoherence from thermal quasiparticles~\cite{Bal2024Apr}\footnote{It is also important to note that both Nb and Ta share a common issue of oxidation (as discussed in the former case by Ref.~\cite{Bal2024Apr}), which is a general experimental problem that must be appropriately managed regardless of e-p coupling or the breakdown of amorphous structure.}. When considering materials with larger electron-phonon coupling, it is therefore better to consider oxides that naturally form an irregular crystal as opposed to an amorphous structure, such as Ta~\cite{Tyner2025}, as they will host fewer low-frequency TLSs. Additionally, the effect on the $T_1$ time from a local amplification of the critical current (at least in the case of a phase qubit) is smaller if we consider larger Josephson junctions. This issue is more difficult to resolve for certain devices, as the Josephson junction size is roughly fixed to ensure an appropriate $\epsilon_J$~\cite{Manenti2023Aug}. While smaller junctions might be favourable in some cases in the sense that they will have fewer TLSs~\cite{Martinis2005Nov}, our work suggests that any remaining TLSs in both the interface and superconducting bulk %of these smaller junctions 
will have a considerable effect on the average critical current if the intrinsic electron phonon-coupling of the material is larger than that of Ta.

\section{VI. Acknowledgments}
%\acknowledgments{} 
We thank Rufus Boyack,  Mikael Fogelström, Ivan Khaymovich, Ali Gokirmak, Josh Mutus, David Pappas, Yaniv Rosen, David Santiago, Jim Sauls,  Irfan Siddiqi, Bill Shelton,  Ilya Sochnikov, Christopher Spitzer, Ilya Vekhter, and Patrick Wong for useful discussions. We also thank Daniel Sandager Dragheim Kjær, Pei Liu, and the entire NNF Quantum Computing Programme for providing SEM pictures of their aluminium Josephson junctions.
This work is supported by the Novo Nordisk Foundation, Grant number NNF22SA0081175, NNF Quantum Computing Programme. AB and JH were also supported by European
Research Council under the European Union Seventh Framework ERS-2018-SYG 810451 HERO and by Knut and Alice Wallenberg Foundation Grant No. KAW
2019.0068.

\bibliography{main}{}
%\newpage 
\newpage
\section*{Appendix A: Derivation of the local density of states}

In this appendix, we will derive Eqn. \eqref{LDOS}, which quantifies the change in the local density of states (LDOS) from inelastic scattering of normal electrons with the localized Einstein impurity. To begin, we recall that the full density of states is given by the imaginary part of the total Green's function:
\begin{align}
    N({\bf r}_i,\,\omega)=-\dfrac{1}{\pi}\Im [G({\bf r}_i,\,{\bf r}_i;\omega)].
\end{align}

\noindent The total Green's function is given by Eqn.~\eqref{Green} in the main text. The change in the density of states is induced by the self energy, which is given in Eqn.~\eqref{self}.

% , is reproduced below for convenience:
% \begin{align}
%     G({\bf r}_{i},\,{\bf r}_{i};\,\omega)&=G^0({\bf r}_i,\,{\bf r}_i;\,\omega)\notag\\
%     &+\sum_{j,\ell}G^{0}({\bf r}_{i},\,{\bf r}_{j};\,\omega)\Sigma({\bf r}_{j},{\bf {\bf r}_{\ell}};\,\omega)G^{0}({\bf r}_{\ell},\,{\bf r}_{i};\,\omega).
% \end{align}

% The change in the density of states is induced by the self energy, which is given by

% \begin{align}
% &\Sigma({\bf r}_{j},\,{\bf r}_\ell \,;\omega)\notag\\
% &=-g^2\int d\omega'\,D^{0}({\bf r}_{j},\,{\bf r}_\ell ;\,\omega-\omega')G^{0}\left({\bf r}_{j},\,{\bf r}_\ell ;\,\omega')\right),
% \end{align}

% \noindent which is Eqn.~\eqref{self} in the main text.

For the purposes of our study, we will calculate the self energy induced by normal electrons scattering off the TLS, the latter of which we are modeling as a localized bosonic mode. To proceed in the calculation, we analytically continue to the imaginary frequency axis; i.e., we let $\omega\equiv \lim_{\delta\rightarrow 0}(\omega+i\delta)\rightarrow i\omega_n$. In this way, we re-define the self energy in terms of the local, finite-temperature Matsubara Green's function $G^{(0)}({\bf r}_j,\,{\bf r}_\ell;\,i\omega_m)$. This transforms the real-frequency integral given in Eqn.~\eqref{Green} into a sum over the fermionic Matsubara frequencies $\omega_m\equiv(2m+1)\pi /\beta$~\cite{AGDBook,MahanBook}, 
\begin{align}
    &\phantom{=}\Sigma({\bf r}_j,\,{\bf r}_\ell;\,i\omega_n)\notag\\
    %&=-\dfrac{g^2}{\beta}\sum_{\omega_n} D^0({\bf r}_j,\,{\bf r}_\ell;i\omega_n-i\omega_m)G^0({\bf r}_j,\,{\bf r}_\ell;i\omega_m)\notag\\
    &=-\dfrac{\lambda \omega_0}{2N(0)\beta}\sum_{m} D^0({\bf r}_j,\,{\bf r}_\ell;i\omega_n-i\omega_m)G^0({\bf r}_j,\,{\bf r}_\ell;i\omega_m),\label{self2}
\end{align}
where $m\in \mathbb{Z}$ is the Matsubara index and $\beta=1/T$ is the inverse temperature. Note that we have explicitly assumed an Einstein phonon model (i.e., where only a single dispersionless phonon mode with frequency $\omega_0$ couples to the electron), and have thus re-written the electron-phonon coupling strength in its dimensionless form $\lambda\equiv 2N(0)g^2/\omega_0$ (see Appendix C for more detailed explanation). As mentioned in the main text, we will take an Einstein phonon model to describe the phonons in the material as well as the bosonic modes describing the TLS, with the former described by a frequency $\omega_E$ and the latter by $\Omega\ll \omega_E$. Such a phonon model is often used within the context of electron-phonon coupled materials due to its simplicity~\cite{Engelsberg1963Aug,Dogan2003Oct,Marsiglio2020Jun}, and allows us to describe the anti-correlation between $\lambda$ and $\omega_0$ already noted in some disordered media~\cite{Lee2006Aug,Balatsky2006Sep,Niestemski2007Dec,Jenkins2009Nov,Fasano2010Oct,Shan2012May}. 

In the above, the non-interacting Green's function in real space is given by $-i\pi N(0)\textrm{sgn}(\omega_m)\mathcal{P}({\bf r}_{j\ell})$, where the spatial modulation factor $\mathcal{P}({\bf r}_{j\ell})$ is given in Eqn.~\eqref{modulation} and follows from the spatial Fourier transform of the fermionic Green's function in a metal~\cite{Weiss1958Aug}. Direct insertion into Eqn.~\eqref{self2} yields
\begin{align}
&\Sigma({\bf r}_{j},{\bf r}_{\ell};i\omega_{n})\notag\\
&=\dfrac{i\pi \lambda \omega_0}{2\beta}\mathcal{P}({\bf r}_{j\ell})\sum_{m}D({\bf r}_{j}-{\bf r}_{\ell};i\omega_{n}-i\omega_{m})\textrm{sgn}(\omega_{m}).\label{self3}
\end{align}
Within the Einstein phonon model, the above sum becomes analytically tractable by invoking the form of the boson propagator on the imaginary frequency axis,
\begin{align}
    D({\bf r}_{j}-{\bf r}_{\ell};i\omega_{n}-i\omega_{m})=-\dfrac{2\omega_{0}}{(\omega_{n}-\omega_{m})^{2}+\omega_{0}^{2}},
\end{align}
in which case the sum given in Eqn.~\eqref{self3} becomes
\begin{align}
    &\phantom{=}\sum_{m=-\infty}^{\infty}\dfrac{\omega_{0}}{(\omega_{n}-\omega_{m})^{2}+\omega_{0}^{2}}\textrm{sgn}(\omega_m)\notag\\
    &=\sum_{m=0}^{\infty}\bigg[\dfrac{\omega_{0}}{(\omega_{n}-\omega_{m})^{2}+\omega_{0}^{2}}-\dfrac{\omega_{0}}{(\omega_{n}+\omega_{m})^{2}+\omega_{0}^{2}}\bigg]\notag\\
    &=\sum_{m=0}^{2n}\dfrac{\omega_{0}}{(\omega_{n}-\omega_{m})^{2}+\omega_{0}^{2}}\notag\\
    &=\dfrac{1}{\omega_{0}}\bigg\{1+2\sum_{m=0}^{n-1}\dfrac{\omega_{0}^{2}}{\nu_{m+1}^{2}+\omega_{0}^{2}}\bigg\}\label{summation},
\end{align}
where in the last line we have used the fact that $\omega_n-\omega_m= \nu_{n-m}$, where $\nu_{n-m}$ is the bosonic Matsubara frequency. To perform the summation, we rewrite the sum given in Eqn.~\eqref{summation} in terms of the hyperbolic tangent and the digamma function $\psi(...)$ by virtue of the identities $\sum_{m=0}^{N}1/(m+x)=\psi(x+N+1)-\psi(x)$ and $\psi(1+ix)-\psi(1-ix)={1}/(ix)-\pi\cot(\pi ix)$:

%\begin{widetext}
\begin{align}
    &2\sum_{m=0}^{n-1}\dfrac{\omega_{0}^{2}}{\nu_{m+1}^{2}+\omega_{0}^{2}}\notag\\
    &\notag\\
    &=\dfrac{\omega_{0}}{2\pi Ti}\left[\sum_{m=0}^{n-1}\dfrac{1}{m+1-\dfrac{i\omega_{0}}{2\pi T}}-\sum_{m=0}^{n-1}\dfrac{1}{m+1+\dfrac{i\omega_{0}}{2\pi T}}\right]\notag\\
    &\notag\\
    &=-1+\dfrac{\omega_{0}}{2T}\coth(\dfrac{\omega_{0}}{2T})\notag\\
    &\phantom{=\,\,\,}-i\dfrac{\omega_{0}}{2\pi T}\left[\psi\left(\dfrac{1}{2}-i\dfrac{\omega_{0}+i\omega_{n}}{2\pi T}\right)-\psi\left(\dfrac{1}{2}+i\dfrac{\omega_{0}-i\omega_{n}}{2\pi T}\right)\right].
\end{align}
%\end{widetext}
A similar expression is derived within the context of pair breaking in anisotropic superconductors~\cite{Millis1988Apr} and Homes scaling in an electron-phonon superconductor~\cite{Heath2024Jul}.

% Final result:
% \begin{align}
% \dfrac{\delta N({\bf r}_{i},\omega)}{N(0)}=\dfrac{\pi^{2}}{2}\Omega_{0}N(0)\sum_{j}\delta_{j\ell}\lambda_j\mathcal{P}^{2}({\bf r}_{ij})\Theta\left(|\overline{\omega}|-1\right)\label{LDOS},
% \end{align}

% \begin{align}
% \Sigma(\omega+i\delta)=    N(0)\lambda^{2}J_{0}(k_{F}(r_{j}-r_{\ell}))\bigg\{ i\pi\coth(\dfrac{\omega_{0}}{2T})+\psi\left(\dfrac{1}{2}-i\dfrac{\omega_{0}+\omega}{2\pi T}\right)-\psi\left(\dfrac{1}{2}+i\dfrac{\omega_{0}-\omega}{2\pi T}\right)\bigg\}
% \end{align}

Taking the limit of $T\rightarrow 0$, the above simplifies greatly, as $\lim_{z\rightarrow\infty}\psi\left({1}/{2}+z\right)\approx\log(z)$ and, as a consequence,\newpage \vspace{0mm}\phantom{.}\vspace{-3.99mm}\\\noindent the difference of the digamma functions is given by $\psi\left[1/2-i(\omega_{0}+\omega)/(2\pi T)\right]-\psi\left[{1}/{2}+i({\omega_{0}-\omega})/(2\pi T)\right]=-i\pi\left[1-\Theta(|\omega|-\omega_{0})\right]$. The expression for the self energy then becomes

\begin{align}
&\Sigma({\bf r}_{j},{\bf r}_{\ell};i\omega_{n})=\dfrac{i\pi \lambda \omega_0}{2}\mathcal{P}({\bf r}_{j\ell})\Theta(|\omega|-\omega_0).
\end{align}

\noindent Assuming a local form for the self energy (i.e., that $\Sigma({\bf r}_j,\,{\bf r}_\ell;,\,i\omega_n)=\delta_{j\ell}\Sigma({\bf r}_j,\,{\bf r}_\ell;\,i\omega_n)$~\cite{Metzner1989Jan,Vollhardt2011Sep}) and using the previously defined form of the Green's function and the density of states, we then readily come to Eqn.~\eqref{LDOS} in the main text. Note that, due to the electron-TLS scattering, the electronic density of states obtains an additional contribution which only occurs when the frequency $\omega$ is above a threshold set by the TLS frequency. This is equivalent to what other studies have seen within the context of inelastic electron tunnelling spectroscopy (IETS)~\cite{Stipe1998Jun,Balatsky2003Dec,Zhu2004Jul,Balatsky2006May,Behm2010Dec}.

% \newpage \vspace{0mm}\phantom{.}\vspace{-5.75}\\ \noindent

\section*{Appendix B: Derivation of the $T_1$ dephasing time for a phase qubit}

In the main text, we discuss the implications a single TLS has on the $T_1$ dephasing time of a phase qubit. In this appendix, we explicitly derive the equation for the $T_1$ time for this particular kind of qubit (i.e., Eqn.~\eqref{T1_eqn}), and synthesize the results given in several previous works~\cite{Devoret,Martinis2002Aug,Martinis2003Mar}.

\begin{figure}
%\subfloat[\label{sfig:testa}]{%
 \hspace{-6mm} \includegraphics[width=1\linewidth]{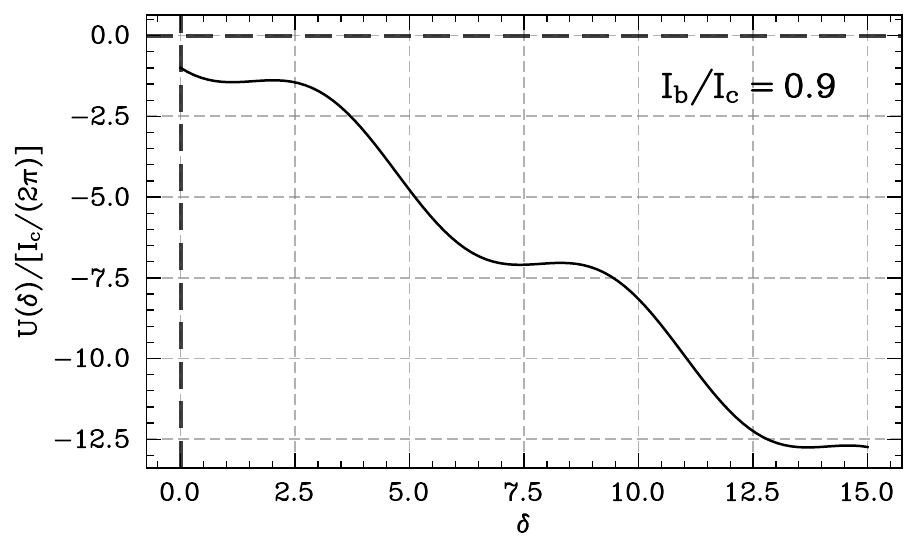}%
%}
\caption{\small Schematic of the washboard potential for a phase qubit, with $I_b/I_c=0.9$. The oscillation frequency about one of the minima may be identified with the qubit frequency under certain assumptions of the bias current. 
\phantom{dfef djfkdsjfk jklsdjkljsf dfjkds;fj dsfj j dsfdsfdsf dsf df fdf sf sd f sd f dsfjsdkf;sj djkfj kdfjs;k sfdjsdfkj dskf jsdklfj klsd}
}
\label{Fig_Washboard}
\end{figure}

The Josephson equations for the current $I(t)$ and voltage $V(t)$ through the Josephson junction are~\cite{Josephson1974Apr}:

\begin{subequations}
\begin{equation}
I(t)=I_{c}\sin\delta(t),\label{SC_1}
\end{equation}
\begin{equation}
V(t)=\dfrac{\Phi_{0}}{2\pi}\dfrac{d\delta(t)}{dt},\label{SC_2}
\end{equation}
\end{subequations}
\newpage \vspace{0mm}\phantom{.}\vspace{-4.01mm}\\\noindent
where $\Phi_{0}=h/2e$ is the flux quantum and $\delta=\phi_{L}-\phi_{R}$ is the phase difference between the right and left superconductors of the junction. As long as the bias current $I_b$ through the device is less than $I_c$, there will be no voltage drop across the junction, and thus $\delta(t)=\delta$ is a constant. The total energy of the system is given in the main text as Eqn.~\eqref{PhaseHam}, and is the sum of the total energy from the effective capacitance $C$, the critical Josephson current $I_c$, and the bias current $I_b$. The oscillation about one of the minima of the resultant "washboard potential" (see Fig.~\ref{Fig_Washboard}) is given by $\omega_{J}=1/\sqrt{L_J C}$, where the effective Josephson inductance $L_J$ follows from Eqn.~\eqref{SC_2}. As the charge $Q$ and phase $\delta$ are conjugate variables, we may quantize Eqn.~\eqref{PhaseHam} to obtain the full quantum Hamiltonian. Expansion about the potential minima results in a Hamiltonian qubit with off-diagonal components controlled by the microwave component to the bias current $\Delta I$. Assuming that the local minimum is sufficiently deep (specifically, $\Delta U\gg 5 \hbar \omega_0/36$~\cite{LandauBook, Martinis2002Aug}), we can approximate $\omega_J\approx \omega_{01}$, where $\omega_{01}$ is the lowest energy level splitting of the qubit. % \begin{align}
% H=\left(\begin{array}{cc}
% \hbar\omega_{0}/2 & -\Delta I\sqrt{{\hbar}/{(2\omega_{0}C)}}\\
% -\Delta I\sqrt{{\hbar}/{(2\omega_{0}C)}} & 3\hbar\omega_{0}/2
% \end{array}\right)\label{Bias_Ham}
% \end{align}
%Assuming that the bias current $I_b$ is slightly less than the critical current, we can use Eqn.~\eqref{SC_1} to rewrite the resonant frequency as $\omega_0=\sqrt{{2\pi I_{c}}/({\Phi_{0}C})}\left[2\left(1-{I_{b}}/{I_{c}}\right)\right]^{1/4}
%$. 
%To implement control of the qubit, we consider 
%Taking $\Delta I=0$, Eqn.~\ref{Bias_Ham} simplifies to $H=\sigma_{0}+\dfrac{1}{2}{\bf d}\cdot{\bf {\bf \sigma}}$ where ${\bf d}=(0\,\,\,0\,\,\,-\hbar \omega_0/2)^T$. This defines a Hamiltonian qubit with Bloch vector given by ${\bf d}$, however in the absence of a bias current the Bloch vector simply rotates about the z-axis. To implement quantum gates, we introduce 
%an additional time-dependence to 
Taking a time-dependent component to the bias current given by $\Delta I(t)=-I_{\mu w,c}(t)\cos(\omega_{01}t)-I_{\mu w,s}(t)\sin(\omega_{01}t)$, the qubit Hamiltonian in the rotating frame may be written as
%In the rotating frame, this yields a qubit Hamiltonian given by
%Eqn.~\eqref{Bias_Ham} may be written as
% \begin{align}
%    H_{rf}(t)\approx \dfrac{1}{2}\sqrt{\dfrac{\hbar}{2\omega_{0}C}}\bigg[I_{\mu w,c}(t)\sigma_{x}+I_{\mu w,s}(t)\sigma_{y}\bigg]
% \end{align}
% Assuming that $I_{\mu w}(t)$ is constant over a short time interval $\Delta t$, the above simplifies to 
$H_{rf}=[{\hbar}/({2\Delta t})] {\bf c}\cdot\sigma$, where we have assumed that $I_{\mu w,\,c}(t)$ and $I_{\mu w,\,s}(t)$ are constant over a short time interval $\Delta t$ and where we have taken ${\bf c}=(\Delta t/\hbar)\sqrt{\hbar/(2\omega_{01} C)} ( I_{\mu w,c}\,\,\,I_{\mu w,s}\,\,\,0)^T$ as a control vector. Time evolution of the Bloch vector in the presence of the bias current is now given by the unitary $U=e^{-i\sigma\cdot{\bf c}/2}$, which defines a rotation about the control axis {\bf c}/|{\bf c}| with angle $\theta\equiv |{\bf c}|/2$ given by $\theta=|{\bf c}|=[{\Delta t}/{\sqrt{2\omega_{01}C\hbar}}]\cdot \sqrt{I_{\mu w,c}^{2}+I_{\mu w,s}^{2}}$. 

\begin{figure}
%\subfloat[\label{sfig:testa}]{%
 \hspace{-6mm} 
 \vspace{15.7mm} \includegraphics[width=1\linewidth]{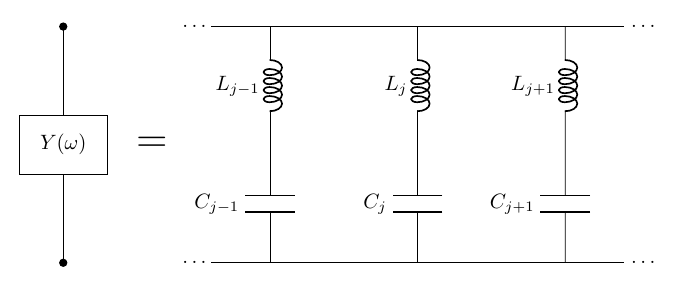}%
%}
\caption{\small Representation of the effective admittance $Y(\omega)$ of the noisy Josephson junction (left) as a collection of LC circuits connected in parallel (right). Figure adapted from Ref.~\cite{Devoret}.
}
\label{Fig_CLCircuit}
\end{figure}

The $T_1$ time may be derived by considering noisy fluctuations in the bias current. Taking some fluctuation of $\Delta I$ to be quantified by either $\delta I_{c}$ or $\delta I_{s}$, the control vector is now modified by ${\bf c}={\bf c}_0+\delta {\bf c}$, and thus the new angle of rotation about the control axis is given by $\theta=\theta_0+\delta \theta$, where $\delta \theta =[\Delta t/(\sqrt{2\omega_{01} C\hbar})]\cdot I_n(t)$ and we define $I_n(t)\equiv \sqrt{(I_{\mu w,c}+\delta I_{c})^{2}+(I_{\mu w,s}+\delta I_{s})^{2}}-\sqrt{I_{\mu w,c}^{2}+I_{\mu w,s}^{2}}$. The total accumulated noise of the angle about the control vector over a time t (which we'll call $\theta_{n}(t)$) is then the integral of $\delta \theta(t)$ in the limit of $\delta t\rightarrow dt$. These small rotations lead to noise in the transitions between the $|0\rangle$ and $|1\rangle$ states of the phase qubit. We formally define noise of the phase qubit itself in terms of the mean-square of the noise in the rotation angle, which we write in terms of a current-current correlator for $I_n(t)$:

\begin{align}
\langle\delta\theta^{2}(t)\rangle&=    \dfrac{1}{2\omega_{01}C\hbar}\int_{0}^{t}dt'\int_{0}^{t}dt''\langle I_{n}(t')I_{n}(t'')\rangle\notag\\
&=\dfrac{1}{2\omega_{01}C\hbar}\int_{-\infty}^{\infty}df\,S_{I}(f)W_{0}(f)\notag\\
&=\dfrac{t}{2\omega_{01}C\hbar}S_{I}(\omega_{01}),\label{RMS}
\end{align}
\noindent where in the second line we have re-written the correlator in terms of the spectral density $S_I(f)$ and the spectral weight function $W_{0}(f)={\sin^{2}(\pi ft)}/{(\pi f)^{2}}$, and in the last line we have approximated the value of the spectral density to be a constant immediately about the noise source at $f=\omega_{01}$. Eqn.~\eqref{RMS} allows us to define the $T_1$ dephasing time as the characteristic time scale associated with error in the transition from state $|0\rangle$ to $|1\rangle$:
\begin{align}
\dfrac{1}{T_{1}}=\dfrac{1}{2\hbar\omega_{01}C}S_{I}(\omega_{01}).
\end{align}

We will now connect the spectral density of the current-current fluctuations to the dissipative properties of the medium via the Caldeira-Leggett approximation~\cite{Devoret}, which models the "lossy" behaviour of the junction as an infinite collection of harmonic oscillators. This solves the problem of the system evolving unitarily while also having dissipation inducing some non-reversible effect; i.e., the energy from the circuit is distributed among a huge number of degrees of freedom, and is unlikely to return to the Josephson junction. We will likewise model this bath of infinite harmonic oscillators as a set of LC circuits in parallel with the Josephson junction (see Fig.~\ref{Fig_CLCircuit}). To ensure that the LC circuits appropriately model the loss induced by the material in the junction, we need to ensure that the real admittance $\Re(Y(\omega))$ of the material matches the admittance of the coupled LC circuits. It can be shown that the admittance of the $j$th LC circuit is given by $\Re(Y_{j}(\omega))=[{\pi}/{(2L_{j})}]\cdot\left[\delta(\omega-\omega_{j})+\delta(\omega+\omega_{j})\right]$ by invoking the Sokhotski-Plemelj formula, and thus we restrict the inductance $L_j$ of the $j$th LC circuit to be $L_{j}={\pi}/[{2\Delta\omega_{j}\Re(Y(\omega_{j}))}]$ and where $\Delta\omega_{j}$ is taken to be the (small) difference between the nearest-neighbour LC branches. From this condition on the individual inductance, we may then relate the spectral density $S_I(\omega_0)$ to the real admittance of the material $\Re(Y(\omega))$, which follows from writing the total current-current correlator in terms of the correlators over each branch of the LC circuit:

\begin{align}
   \langle I(t)I(0)\rangle&=\sum_j \langle \dfrac{\phi_j(t)}{L_j}\dfrac{\phi_j(0)}{L_j}\rangle\notag\\ 
   &=\sum_{j}\dfrac{\hbar\omega_{j}}{2L_{j}}\left(n(\omega_{j})e^{i\omega_{j}t}+[n(\omega_{j})+1]e^{-i\omega_{j}t}\right)\notag\\
    &\equiv \sum_j \dfrac{\hbar \omega_j}{2L_j} \mathcal{N}(\omega_{j},\,t),\label{II}
\end{align}
where $\phi_j(t)$ is the flux of the $j$th LC circuit and in the penultimate line we have summed over all branches and we have defined $n(\omega_{j})\equiv \langle a_{j}^{\dagger}a_{j}\rangle$ in terms of the bosonic operators $a_m$ and $a_j^\dagger$ of the quantized LC circuit. Taking $L_{j}={\pi}/[2\Delta\omega_{j}\Re(Y(\omega_{j}))]$ and the limit $\Delta\omega_{j}\rightarrow d\omega_j$, the current-current spectral density follows from Eqn.~\eqref{II}:

\begin{align}
    S_{I}(\omega)&=\dfrac{\hbar}{\pi}\int_{-\infty}^{\infty}d\omega_{j}\omega_{j}\Re(Y(\omega_{j}))\int_{-\infty}^{\infty}dt\,\mathcal{N}(\omega_{j},\,t)e^{i\omega t}\notag\\
    &=\dfrac{2\hbar\omega\Re(Y(\omega))}{1-e^{-\beta\hbar\omega}}.
\end{align}

\noindent The zero-point noise is then given by $S_{I}(\omega_{01})-S_{I}(-\omega_{01})=2\hbar\omega_{01}\Re(Y(\omega_{01}))$, and the $T_1$ time is then simplified to

\begin{align}
 \dfrac{1}{T_1}=   \dfrac{\Re(Y(\omega_{01}))}{C}.\label{T1_first}
\end{align}

To further simplify, we will assume the admittance is dominated by the effective capacitance, and thus $Y(\omega)\approx-i\omega C(\omega)$. Likewise, splitting the capacitance into a real and imaginary (i.e., "lossy") component, the admittance is further simplified to $\Re [Y(\omega)]\approx \omega\Im[C(\omega)]$. Finally, we define the loss tangent as the ratio (or tangent of the angle) between the imaginary and real permittivities; i.e., $\tan\delta =\Im(\epsilon)/\Re(\epsilon)$. As the capacitance of the junction is directly proportional to the permittivity, we may write $\Im(C)=\alpha C\tan \delta$, where we have also taken $\Re(C)=\alpha C$ where $\alpha$ is some number less than one (meaning that there is an increase in energy loss within the capacitor due to dissipation). This results in the real admittance $\Re(Y(\omega_{01}))=\omega_{01}\alpha C\tan\delta$. Plugging this expression into Eqn.~\eqref{T1_first}, we arrive at Eqn.~\eqref{T1_eqn} given in the main text.
%\phantom{dfadfd dsf fjwkjdf k ls;kjf k;ljsdfkl jk;s jfdksjfdklj kjk fsjk j df d dsf ds df sf f}
%\\\\
\section*{Appendix C: Technical background on Eliashberg theory and numerical details}

In order to appropriately quantify the influence of a localized Einstein boson on the Josephson critical current, we consider the full frequency-dependent (i.e., dynamical) gap function within Eliashberg theory~\cite{Eliashberg1,Eliashberg2}. In Eliashberg theory, the Cooper pairing of electrons is described in a way that takes the dynamical nature of the phonon into account. In this way, Eliashberg theory may be seen as a non-trivial extension of BCS theory, where in the latter case the gap is considered to be a static quantity. While there are many excellent reviews and resources for learning Eliashberg theory~\cite{RickayzenBook, Parks1, AlexandrovBook, Bennemann, ummarino, Chubukov2020Jun, Protter2021}, in this appendix we will limit ourselves to an explanation of why incorporating dynamical phonon effects is crucial for studying superconducting materials applicable to contemporary quantum technology.

The full frequency-dependent gap function may be found via the self-consistent Eliashberg equations on the imaginary (i.e., Matsubara) frequency axis~\cite{Bennemann,Marsiglio2020Jun}:
\begin{widetext}
\begin{subequations}
\begin{equation}
    \phi({\bf k},\,i\omega_n)=\dfrac{1}{\beta N}\sum_{{\bf k}',\,m}\left[ |g_{{\bf k}-{\bf k}'}|^2 D({\bf k}-{\bf k}',i\omega_n-i\omega_{m})-V_{{\bf k}{\bf k}'}\right]\dfrac{
\phi({\bf k}',\,i\omega_m)
    }{\widetilde{\omega}^2(i\omega_m)+[\xi_{{\bf k}'}+\chi({\bf k}',i\omega_m)]^2+\phi^2({\bf k}',i\omega_{m})},\label{ETh1}
\end{equation}
\begin{equation}
    \widetilde{\omega}(i\omega_n)-\omega_n=\dfrac{1}{\beta N}\sum_{{\bf k}',\,m} |g_{{\bf k}-{\bf k}'}|^2D({\bf k}-{\bf k}',i\omega_n-i\omega_{m})\dfrac{
\widetilde{\omega}(i\omega_m)
    }{\widetilde{\omega}^2(i\omega_m)+[\xi_{{\bf k}'}+\chi({\bf k}',i\omega_m)]^2+\phi^2({\bf k}',i\omega_{m})},\label{ETh2}
\end{equation}
\begin{equation}
    \chi({\bf k},\,i\omega_n)=-\dfrac{1}{\beta N}\sum_{{\bf k}',\,m} |g_{{\bf k}-{\bf k}'}|^2D({\bf k}-{\bf k}',i\omega_n-i\omega_{m})\dfrac{
\xi_{{\bf k}'}+\chi({\bf k}',i\omega_m)
    }{\widetilde{\omega}^2(i\omega_m)+[\xi_{{\bf k}'}+\chi({\bf k}',i\omega_m)]^2+\phi^2({\bf k}',i\omega_{m})},\label{ETh3}
    \end{equation}
    \begin{equation}
    n=1-\dfrac{2}{\beta N}\sum_{{\bf k}',\,m} |g_{{\bf k}-{\bf k}'}|^2\dfrac{
\xi_{{\bf k}'}+\chi({\bf k}',i\omega_m)
    }{\widetilde{\omega}^2(i\omega_m)+[\xi_{{\bf k}'}+\chi({\bf k}',i\omega_m)]^2+\phi^2({\bf k}',i\omega_{m})}.\label{ETh4}
\end{equation}
\end{subequations}
\end{widetext}
A full derivation of Eqns.~\eqref{ETh1}-\eqref{ETh4} is provided in the introductory book of Rickayzen~\cite{RickayzenBook} as well as in the short review by Marsiglio~\cite{Marsiglio2018Jul}. In the above, $D({\bf k}-{\bf k}',\,i\omega_n-i\omega_m)$ is the bosonic propagator, $g_{{\bf k}-{\bf k}'}$ is the bare electron-phonon coupling strength, $V_{{\bf k}{\bf k}'}$ is a direct Coulomb repulsion between two electrons, and $\xi_{\bf k}=\epsilon_{\bf k}-\mu$ is the shifted electron dispersion. These equations self-consistently describe, respectively, the anomalous component of the self energy (pairing function) $\phi({\bf k},\,i\omega_n)$, the renormalised frequency $\widetilde{\omega}({\bf k},\,i\omega_n)$, the energy shift $\chi({\bf k},\,i\omega_n)$, and the electron number density $n$. The former three quantities are more fundamentally defined in terms of the normal self energy $\Sigma({\bf k},\,i\omega_n)$ and the anomalous (superconducting) Green's function $F({\bf k},\,i\omega_n)$ as follows:
\begin{subequations}
    \begin{align}
        &\phi({\bf k},\,i\omega_n)\notag\\
        &=-\dfrac{1}{\beta N}\sum_{{\bf k}',\,m}|g_{{\bf k}-{\bf k}'}|^2 D({\bf k}-{\bf k}',\,i\omega_n-i\omega_m)F({\bf k}',\,i\omega_m),
    \end{align}
    \begin{equation}
        \widetilde{\omega}_n({\bf k},\,i\omega_n)-\omega_n=\dfrac{i}{2}\left[\Sigma({\bf k},\,i\omega_n)-\Sigma({\bf k},\,-i\omega_n)\right],
    \end{equation}
        \begin{equation}
        \chi({\bf k},\,i\omega_n)=\dfrac{1}{2}\left[\Sigma({\bf k},\,i\omega_n)+\Sigma({\bf k},\,-i\omega_n)\right].
    \end{equation}
\end{subequations}
\noindent The main functions of interest one usually obtains from solving the Eliashberg equations are the renormalisation function $Z({\bf k},\,i\omega_n)\equiv \widetilde{\omega}({\bf k},\,i\omega_n)/\omega_n$ and the gap function $\Delta({\bf k},\,i\omega_n)\equiv \phi({\bf k},\,i\omega_n)/Z({\bf k},\,i\omega_n)$. The gap function roughly takes the role of the BCS gap in Eliashberg theory (expanded upon more formally at the end of this appendix), while the renormalisation function quantifies the mass renormalisation from electron-phonon effects. In the normal state, $\Delta({\bf k},\,i\omega_n)=0$ by definition, while $Z({\bf k},\,i\omega_n)$ might still exhibit some non-trivial frequency dependence from the electron-phonon interaction.

The electron-phonon coupling strength is typically quantified in terms of the dimensionless parameter $\lambda_{{\bf k}{\bf k}'}$, given by

\begin{align}
    \lambda_{{\bf k}{\bf k}'}=2\int_0^\infty d\nu \dfrac{\alpha_{{\bf k}{\bf k}'}^2 F(\nu)}{\nu},
\end{align}

\noindent where $\alpha_{{\bf k}{\bf k}'}^{2}F(\nu)=N(0)|g_{{\bf k}-{\bf k}'}|^{2}B(\nu)$ is the electron-phonon spectral function and $B(\nu)$ is the phonon spectral function. In addition to the constraint of $\omega_E/\epsilon_F\ll 1$ (which allows us to ignore vertex corrections in the pairing interaction, neglect Landau damping of the phonons, and permits us to linearize the electronic dispersion near the Fermi energy~\cite{Marsiglio1991Sep,Chubukov2020Jun,Zhang2022Oct}), there remains some debate as to the validity of Eliashberg theory for materials with an electron-phonon coupling strength greater than unity~\cite{Allen1975Aug,Marsiglio1988b,Marsiglio1991Sep,Combescot1995May,PhysRevB.106.064502,Yuzbashyan2022Jul,PhysRevB.106.054518, Esterlis2019}. As we only consider those materials with a bulk electron-phonon coupling strength $\lambda^{(0)}\lessapprox 1$, such an issue is not apposite to this paper. Similarly, note that the complementary weak Eliashberg limit defined by a very small electron-phonon coupling strength {\it should not be considered to be a return to standard BCS theory}~\cite{Marsiglio2018Jul,Boyack2023Mar}. Instead, the traditional BCS gap equation is obtained in the limit of a frequency-independent gap function and a renormalisation function $Z({\bf k},\,i\omega_n)$ of unity. Physically, the former occurs if the frequency-dependent interaction were instantaneous (i.e., if the bosonic propagator is a constant with a hard cutoff)~\cite{RickayzenBook}. Within the full Eliashberg solution, a realistic bosonic propagator results in a frequency-dependent gap function for all material-relevant values of the electron-phonon coupling. For this reason, in our work we study the full gap equation within Eliashberg theory as opposed to solving the BCS gap equation, even when considering more weakly-coupled materials such as aluminium.

We can greatly simplify the Eliashberg equations presented earlier if we assume particle-hole symmetry, a constant electronic density of states, and disregard the direct Coulomb interaction~\cite{Marsiglio2020Jun}. For the sake of considering a general material with arbitrary electron-phonon coupling, we will similarly ignore multiband effects and evaluate the Eliashberg equations at the Fermi level. Then, the above equations reduce to the more digestible single-band Eliashberg equations:

\begin{subequations}
\begin{align}
&\phi(i\omega_{n})= \dfrac{\pi}{\beta}\sum_{m=-\infty}^{\infty}
\frac{\lambda(i\omega_{n}-i\omega_m)\phi({i\omega_m})}{\sqrt{\widetilde{\omega}^2(i\omega_m) + \phi^2({i\omega_m})}}, 
  \label{eq:EthEqnsIm1}\\\notag\\
&Z(i\omega_{n})=  1 + \frac{\pi}{\omega_{n}\beta}\sum_{m=-\infty}^{\infty} 
 \frac{\lambda(i\omega_{n}-i\omega_m)\widetilde{\omega}(i\omega_m)}{\sqrt{\widetilde{\omega}^2(i\omega_m)  + \phi^2({i\omega_m})} },
\label{eq:EthEqnsIm2}
\end{align}
\end{subequations}

\noindent where the Eliashberg function $\lambda(z)$ is given by
\begin{align}
\lambda(z)&=\int_0^\infty d\nu\,\dfrac{2\nu \alpha^2F(\nu)}{\nu^2-z^2}=\dfrac{\lambda^{(0)} \omega_E^2}{\omega_E^2-z^2},
\end{align}
with the final equality being specific to the Einstein phonon model and $\lambda_{{\bf k}{\bf k}'}\Rightarrow\lambda^{(0)}$ now independent of wave vector. Note that, for the equations given above, the values of $\phi(i\omega_n)$ and $Z(i\omega_n)$ are for the clean system. In the presence of some impurity or defect, the scattering rate $\Gamma$ off this impurity must be considered, in which case the pairing function and renormalised frequency become~\cite{Parks2,Marsiglio1992May}

\begin{subequations}
\begin{equation}
\phi_{\textrm{imp}}(i\omega_{n}) = \phi(i\omega_{n})\pm\Gamma\dfrac{\phi(i\omega_{n})}{\sqrt{\widetilde{\omega}^2(i\omega_{n})+\phi^{2}(i\omega_{n})}},\label{EThShort1}
\end{equation}
\begin{equation}
\widetilde{\omega}_{\textrm{imp}}(i\omega_{n}) = \widetilde{\omega}(i\omega_{n})+\Gamma\dfrac{\widetilde{\omega}(i\omega_{n})}{\sqrt{\widetilde{\omega}^2(i\omega_{n})+\phi^{2}(i\omega_{n})}}.\label{EThShort2}
\end{equation}
\end{subequations}
\noindent In Eqns.~\eqref{EThShort1} and \eqref{EThShort2}, the subscript "imp" denotes the pairing function or renormalised frequency in the presence of impurities, while the lack of a subscript denotes the same quantities in the clean limit. Additionally, note that the top (plus) sign in Eqn.~\eqref{EThShort1} denotes the case of a non-magnetic impurity, while the bottom (minus) sign denotes the case of a paramagnetic impurity (i.e., an impurity with a non-zero magnetic dipole moment). For the former scenario, it is easy to see that $\Delta_{\textrm{imp}}(i\omega_n)=\Delta(i\omega_n)$, and thus non-magnetic impurities will not affect the gap function within Eliashberg theory, as required by Anderson's theorem~\cite{Anderson1959Sep,Balatsky2006May}. However, if we assume a paramagnetic impurity, then the gap itself has some dependence upon $\Gamma$. For the Einstein phonon model, one may calculate the scattering rate from the self energy $\Sigma(i\omega_n)$ on the imaginary frequency axis and analytically continue to the real axis to find $\Gamma({\bf r}_i,\omega)=\mathcal{P}({\bf r}_i)\omega \Im(Z(\omega))$. This follows from the form of the normal electrons' self energy (Eqn.~\eqref{self2}) and Eqn.~\eqref{eq:EthEqnsIm2}. Note that the scattering rate has some additional frequency dependence, and thus the localized Einstein boson we use to model an individual TLS may be thought of as a dynamical impurity. Also note that our result for the scattering rate may be derived (sans the spatial dependence) by taking the normal state limit of the DC conductivity in the presence of electron-phonon interactions, and identifying a transport scattering time beyond the Drude type~\cite{Heath2024Jul}.

To solve the Eliashberg equations on the imaginary frequency axis and obtain the clean gap equations, we numerically solve Eqns.~\eqref{eq:EthEqnsIm1} and \eqref{eq:EthEqnsIm2} with a careful self-consistent procedure. If certain convergence criteria are not met for a given upper limit to the Matsubara frequency summation, then the upper limit of the sum is increased until convergence is found. Upon convergence, the Matsubara sum is extended to a maximum index $M\sim 10^4-10^5$ via a weak Eliashberg fit. For further details of the "Ouroboros algorithm" we use, see a previous paper by one of the authors (JTH) and Rufus Boyack~\cite{Heath2024Jul}. In the present manuscript, convergence of the Matsubara sum for each value of $\lambda^{(0)}$ is checked by ensuring that the taxicab (i.e., rectilinear) distance between the $M$ and $M+1$ solutions of the gap function is within $0.15$. Convergence of the larger Matsubara frequency solutions (i.e., the "tail" of the gap function) is guaranteed by confirming that the last $200$ values of the gap function we find iteratively exhibits an $r^2$ value of $0.95$ with the weak Eliashberg prediction. Once the appropriate maximal index $M$ is found for a given $\lambda^{(0)}$, the self-consistent iteration of Eqns.~\eqref{eq:EthEqnsIm1} and ~\eqref{eq:EthEqnsIm2} is performed until an accuracy within $10^{-6}$ is achieved. 

Finally, to find the Josephson current, we analytically continue Eqns.~~\eqref{eq:EthEqnsIm1} and \eqref{eq:EthEqnsIm2} to the real frequency axis so that we may find $\Delta(\omega)$ and $Z(\omega)$. Once again, we use the shorthand notation $\omega\equiv \omega+i\delta$, with $\delta\ll 1$. In older papers, such analytic continuation was performed via N-point Pad\'e approximants~\cite{Vidberg1977Nov}, however this method is imprecise for high-temperatures and frequencies~\cite{Blaschke1982Jun,Leavens1985Jan}. For this reason, we instead utilize the Marsiglio-Schossmann-Carbotte (MSC) method, which directly uses the imaginary axis equations to perform an {\it exact} analytic continuation to the real axis~\cite{Marsiglio1988Apr}. Within this scheme, the real axis pairing function is given by $\phi(\omega)=\phi_{\textrm{inhom}}(\omega)+\phi_{\textrm{hom}}(\omega)$, where 
\begin{align}
\phi_{\textrm{inhom}}(\omega) = \dfrac{\pi}{\beta}\sum_{m=-\infty}^{\infty}
\frac{\lambda(\omega-i\omega_m)\phi({i\omega_m})}{\sqrt{\widetilde{\omega}^2(i\omega_m) + \phi^2({i\omega_m})}}\label{RealETh1}
\end{align}
\noindent corresponds to the the inhomogeneous terms of the gap function on the real axis and
\begin{align}
\phi_{\textrm{hom}}(\omega)&=\frac{i\pi}{2} \lambda\omega_E  \Biggl\{ 
  \frac{\left[n_b(\omega_E)+n_f(\omega_E-\omega)\right]\cdot \phi(\omega-\omega_E )}{\sqrt{[\widetilde{\omega}(\omega) - \widetilde{\omega}(\omega_E) ]^2 - \phi^2(\omega-\omega_E )}}\notag\\
  &\phantom{=\dfrac{i\pi}{2}\lambda \omega_E}+
\frac{\left[n_b(\omega_E)+n_f(\omega_E+\omega)\right]\cdot \phi(\omega+\omega_E )}{\sqrt{[\widetilde{\omega}(\omega) + \widetilde{\omega}(\omega_E)]^2  - \phi^2(\omega+\omega_E )}}\Biggr\}\label{RealETh2}
\end{align}
\noindent corresponds to the homogeneous contribution, and where it follows from our notation that $\widetilde{\omega}(\omega)\equiv \omega Z(\omega)$. A similar expression follows for the renormalisation function $Z(\omega)=Z_{\textrm{inhom}}(\omega)+Z_{\textrm{hom}}(\omega)$, where 
\begin{align}
Z_{\textrm{inhom}}(\omega) = 1+\dfrac{i\pi}{\omega\beta}\sum_{m=-\infty}^{\infty}
\frac{\lambda(\omega-i\omega_m)\widetilde{\omega}({i\omega_m})}{\sqrt{\widetilde{\omega}^2(i\omega_m) + \phi^2({i\omega_m})}},\label{RealETh3}
\end{align}
\begin{align}
  Z_{\textrm{hom}}(\omega)&=\frac{i\pi}{2\omega} \lambda\omega_E  \Biggl\{ 
  \frac{\left[n_b(\omega_E)+n_f(\omega_E-\omega)\right]\cdot \widetilde{\omega}(\omega-\omega_E )}{\sqrt{[\widetilde{\omega}(\omega) - \widetilde{\omega}(\omega_E) ]^2 - \phi^2(\omega-\omega_E )}}\notag\\
  &\phantom{=\dfrac{i\pi}{2}\lambda \omega_E}+
\frac{\left[n_b(\omega_E)+n_f(\omega_E+\omega)\right]\cdot \widetilde{\omega}(\omega+\omega_E )}{\sqrt{[\widetilde{\omega}(\omega) + \widetilde{\omega}(\omega_E)]^2  - \phi^2(\omega+\omega_E )}}\Biggr\}.\label{RealETh4}
\end{align}
Numerically, $\Delta(\omega)$ and $Z(\omega)$ are found by a self-consistent algorithm, as was the case on the imaginary frequency axis. Interestingly, by taking $\omega\equiv \omega+i\delta \rightarrow i\omega_n$ in Eqns. \eqref{RealETh1}--\eqref{RealETh4} and noting that $n_b(\omega_E)+n_F(\omega_E-i\omega_n)=n_b(\omega_E)+n_F(\omega_E+i\omega_n)=0$, we re-obtain Eqns.~\eqref{eq:EthEqnsIm1} and \eqref{eq:EthEqnsIm2}. However, it is worth mentioning that the converse is not true; i.e., one cannot simply take Eqns.~\eqref{eq:EthEqnsIm1} and \eqref{eq:EthEqnsIm2} on the imaginary axis and change $i\omega_n$ to $\omega$ to obtain Eqns. \eqref{RealETh1}--\eqref{RealETh4}. Additionally, note that both $\Delta(\omega)$ and $Z(\omega)$ are in general complex functions. This is in contrast to the imaginary axis solutions $\Delta(i\omega_n)$ and $Z(i\omega_n)$, which are always real. For this reason, it is often convenient to define the purely real "gap edge" at low frequency, defined as $\Delta_1\equiv \Re[\Delta(\omega=\Delta_1)]$~\cite{Fibich1965Apr,Boyack2023Mar}. In tandem with its imaginary counterpart $\Delta_2\equiv \Im[\Delta(\omega=\Delta_1)]$, in certain scenarios we may approximate $\Delta(\omega)\approx \Delta=\Delta_1+i\Delta_2$~\cite{Boyack2023Mar}. In BCS theory, the gap edge reduces to the BCS gap, with $\Delta_2=0$. We may likewise define the equivalent entities for the renormalisation function, with $Z_1\equiv \Re[Z(\omega=\Delta_1)]$ and $Z_2\equiv \Im[Z(\omega=\Delta_1)]$ and where, in the BCS limit, $Z_1=1$ and $Z_2=0$.

\end{document}